%% file: Amo_SST_Collective_ARXIV.tex
\documentclass{iopart}
\usepackage{iopams}  
\usepackage{color}
\usepackage{harvard}
\usepackage{graphicx}
\begin{document}

\title{Collective dynamics of excitons and polaritons in semiconductor nanostructures}

\author{A Amo\footnote{Present address:
Laboratoire Kastler Brossel, Universit\'{e} Pierre et Marie Curie, Ecole Normale Sup\'{e}rieure et CNRS, UPMC Case 74,
4 place Jussieu, 75252 Paris Cedex 05, France},
D Sanvitto and L Vi\~{n}a}

\address{Departamento de F\'{i}sica de Materiales, Universidad Auton\'{o}ma de Madrid, 28049 Madrid, Spain}
\ead{alberto.amo@spectro.jussieu.fr}
\begin{abstract}
Time resolved photoluminescence is a powerful technique to study the collective dynamics of excitons and polaritons
in semiconductor nanostructures. We present a two excitation pulses technique to induce the ultrafast and controlled
quenching of the exciton emission in a quantum well. The depth of the dip is given by the magnitude of the warming
of the carriers induced by the arrival of a laser pulse when an exciton population is already present in the sample.
We use this technique to study the relaxation mechanisms of polaritons in semiconductor microcavities, which are of
great importance to enhance the conditions for their condensation under non-resonant excitation. We also explore
the dynamics of polariton fluids resonantly created in the lower polariton branch in a triggered optical parametric
oscillator configuration, showing evidence of polariton superfluidity, and opening up the way to the real-time
study of quantum fluids.
\end{abstract}

%\maketitle

\input{introduction}
\input{twopulsesQW}
\input{polaritonsLinear}
\input{fluiddynamics}
\input{conclusions}

\section*{Acknowledgments}
This work was performed in collaboration with D.~Ballarini, M.~D.~Mart\'{i}n, L. Klopotowski, F.~P.~Laussy, E.~del~Valle, C.~Tejedor, E.~Kozhemiakina, A.~I.~Toropov, K.~S.~Zhuravlev, D.~Krizhanovskii, M.~S.~Skolnick, D.~Bajoni and J.~Bloch. This work would have not been possible without the high quality samples designed and grown by A.~Lema\^itre, J.~Bloch, J.~S.~Roberts and M.~S.~Skolnick. We thank the Spanish MEC (MAT2008-01555/NAN and QOIT-CSD2006-00019), CAM (S-0505/ESP-0200), and the IMDEA Nanociencia for funding. AA acknowledges financial support for the realization of the PhD thesis from the FPU program of the Spanish Ministry of Science and Education, D S the Ram\'{o}n y Cajal program.

\section*{References}
%\begin{thebibliography}{10}
%\bibliography{articulosalberto}
%%\bibitem{book1} Goosens M, Rahtz S and Mittelbach F 1997 {\it The \LaTeX\ Graphics Companion\/} 
%%(Reading, MA: Addison-Wesley)
%%\bibitem{eps} Reckdahl K 1997 {\it Using Imported Graphics in \LaTeX\ } (search CTAN for the file `epslatex.pdf')

\end{document}

%% file: introduction.tex
\section{Introduction}

Semiconductor nanostructures offer a privileged workbench for the study of many fundamental properties of the light-matter interaction and of the collective excitations in solids. Due to single atomic monolayer resolution achieved with epitaxial growth techniques, semiconductor devices can be designed into heterostructures in which the dimensionality of the excitations, the strength of the light-matter interaction and the particle character according to its statistics (bosonic or fermionic) can be finely controlled. Additionally, if the materials of choice in the structure present a direct gap, excitations in its basic form of electrons promoted from the valence to the conduction band can be easily created and detected by optical means.

In quantum wells (QWs), optical excitation leads to the formation of two types of populations: free electrons and holes, and exciton complexes. The two types of populations coexist in quasi thermal equilibrium~\cite{Szczytko2004,Chatterjee2004,Bajoni2006b}, with a temperature which decreases in time towards the lattice temperature when the excitation is pulsed~\cite{vonderLinde1979,Capizzi1984,Leo1988,Yoon1996,Bajoni2006b}. By varying the lattice temperature and density of the photoexcited carriers it is possible to control the ratio between the two populations, allowing for the observation of a transition from an exciton dominated phase (insulating due to the neutral character of these quasiparticles) to a free carrier phase (i. e. conducting)~\cite{Kaindl2003,Kappei2005,Amo2006,Stern2008}. Additionally, the energy separation of the exciton and free carrier recombination enables the detailed study of phase-space filling effects associated to the fermionic character of the free electron-hole populations~\cite{Kappei2005}. For instance, Pauli blockade is one of the typical effects in a fermionic degenerate system~\cite{Warburton1997,Kalevich2001,Ono2002}, and has been shown to greatly alter the electron spin-flip dynamics in semiconductors and, consequently, the polarisation dynamics of the light emitted by the system~\cite{Potemski1999,Dzhioev2002,Nemec2005,Amo2007b}.

Semiconductor nanostructures allow also for the study of the many body properties of boson ensembles. In particular, microcavities consitute an excellent playground. In these systems the fundamental excitations are polaritons: bosons formed from the linear combination of quantum well excitons embedded in a cavity, and the photon modes confined by Bragg mirrors. Due to their partially photonic nature, polaritons have a very small mass, ($\sim10^{-5}~m_e$, the free electron mass) and, consequently, a very high critical temperature for Bose-Einstein condensation (BEC)~\cite{Kasprzak2006,Christopoulos2007,Christmann2008}. Additionally, the properties of the ensemble can be easily probed through the light escaping from the cavity, which arises from the annihilation of polaritons and contains all the energy, coherence and density information of the polariton ensemble inside the cavity. Recent experiments have shown the achievement of polariton condensates in CdTe and GaAs based microcavities at temperatures of the order of $\sim$10~K~\cite{Kasprzak2006,Balili2007,Wertz2009}. Despite their out of equilibrium character (their lifetimes range up to $\sim$10~ps) polaritons have shown similar phenomenology to that observed in atomic condensates. The spontaneous~\cite{Lagoudakis2008} and imprinted~\cite{Sanvitto2009} appearance of quantized vortices, long-range order~\cite{Lai2007b} or Bogoliubov like spectra of excitations~\cite{Utsunomiya2008} are among some of the phenomena associated to a condensate of interacting bosons that have been reported in the microcavity system. However, the observation of this phenomenology in a semiconductor microcavity does not depart significantly from what can be observed in purely photonic systems, such as in vertical cavity surface emitting lasers~\cite{Scheuer1999,Bajoni2007}. A milestone in the study of the macroscopic quantum character of interacting boson systems in semiconductors would be the observation of superfluidity, understood in its most general sense: the absence of friction of a particle ensemble when traversing an obstacle~\cite{Amo2009,Amo2008}.

In this tutorial we present several optical approaches to manipulate the distribution of excited particles in two types of semiconductor nanostructures. We will do this by using different experimental configurations in an all optical set up. In section~\ref{twopulsesQW} we focus on
% a fermionic system: 
photoexcited electrons in QWs. We will see that the photoluminescence spectra are characterized by exciton and free electron-hole recombination. Free carriers in the density regime under study can be well described by a Boltzmann distribution whose temperature can be extracted from the spectral characteristics of the free electron-hole recombination. We will show, that the temperature of pre-phototexcited carriers can be altered in the picosecond timescale by the arrival of a short pulse. The induced ultrafast warming of the electron populations results in an abrupt switch-off of the exciton emission.

In sections~\ref{polaritonslinear} and~\ref{fluiddynamics} we concentrate on a bosonic system: polaritons in semiconductor microcavities. Contrary to electrons, bosons can macroscopically condense in a quantum state, at sufficiently low temperatures. The most extended configuration employed in microcavities to explore this phenomenon is the use of the energy trap in momentum space found in the lower polariton branch dispersion around $k=0$. Under non-resonant excitation, polaritons are formed in the \emph{reservoir} at high momenta, outside the trap. During their relaxation towards lower energy states, due to their particular dispersion which prevents the efficient relaxation of energy and momentum, polaritons tend to accumulate just outside the trap in the so called bottleneck region~\cite{Tartakovskii2000,Senellart2000}.

Relaxation from the bottleneck region to the bottom of the polariton
trap is a slow process at low excitation densities, as it requires
the simultaneous relaxation of significant amounts of energy and momentum~\cite{Malpuech2002}. This so-called bottleneck effect has prevented the observation of polariton condensation in GaAs based microcavities until very recently~\cite{Balili2007,Wertz2009}.
%LO-phonon assisted scattering processes cannot participate due to
%energy conservation constraints, and only polariton-polariton and
%acoustic phonon-polariton interaction can mediate in the polariton
%relaxation.
In section~\ref{polaritonslinear} we will show experimental results on the transition from incoherent polariton emission to photon lasing when increasing the excitation density in a GaAs based microcavity, as a consequence of the bottleneck effect. Additionally, we will show results on the ultrafast modification of the distribution of polaritons in the bottleneck under two pulses photoexcitation, similar to the configuration presented in section~\ref{twopulsesQW} for QWs. These experiments demonstrate that the ultrafast warming of polariton gases is possible, with potential applications on the study of the polariton condensation and on the overcoming of the bottleneck problem in the polariton relaxation~\cite{Tartakovskii2000,Krizhanovskii2004,Perrin2005}.

Finally, in section~\ref{fluiddynamics} we address the dynamic properties of polariton condensates resonantly created on the lower polariton branch. Using a novel excitation configuration based on the optical parametric oscillator, we can create polariton wavepackets with a precise non-zero momentum and a lifetime three orders of magnitude larger than that of polaritons~\cite{Amo2009}. Additionally, by use of a time-resolved high-resolution imaging set-up, we can follow the real- and momentum-space time evolution of the polariton wavepackets. We observe a strong linearization of the dispersion and we find evidence of superfluid behaviour of polaritons. The introduced experimental configuration opens up the way to the study of a variety of collective effects, like the formation of solitons or vortices, and the study of non-spontaneous phase transitions.

%% file: twopulsesQW.tex
\section{Ultrafast warming of carriers in QWs}
\label{twopulsesQW}

Our first approach to the ultrafast optical manipulation of carrier distributions will be developed in semiconductor QWs. 
After a pulsed photoexcitation of a QW, the photoluminescence (PL) dynamics are dominated by the exciton emission. The time evolution of the emitted intensity is determined by the interplay of exciton formation dynamics, its recombination time and by the thermalization and cooling of the free electrons and holes from which the exciton populations are fed \cite{Kaindl2003,Szczytko2004,Chatterjee2004,Amo2006}. Once free carriers have thermalized after the first few hundreds of femtoseconds \cite{Rota1993b,Alexandrou1995}, the strong Coulomb-dipole interaction forces the exciton population to stay in thermal quasiequilibrium with the bath of electrons and holes \cite{Szczytko2005,Bajoni2006b}. This means that the temperature and, therefore, momentum distribution of both species, are the same at all times. As the exciton PL arises just from states with almost no in-plane momentum ($k\sim 0$) due to momentum conservation constraints during the recombination process, any change of temperature in the exciton population will be manifested in the exciton emission. In this section we present experimental results on the optical manipulation of a photogenerated exciton population
in QWs, by applying a delayed, non-resonant optical pulse. This pulse abruptly warms an already-thermalized free electron-hole plasma in the QW which, consequently, causes a sudden warming of the exciton populations. The direct consequence of the abrupt heating of the exciton distribution is the appearance of a sharp dip in PL emission when the delayed pulse reaches the sample \cite{Amo2008b}.

\subsection{Photoluminescence under two pulses excitation}

\begin{figure}
\centering\includegraphics[clip,width=0.9\textwidth]{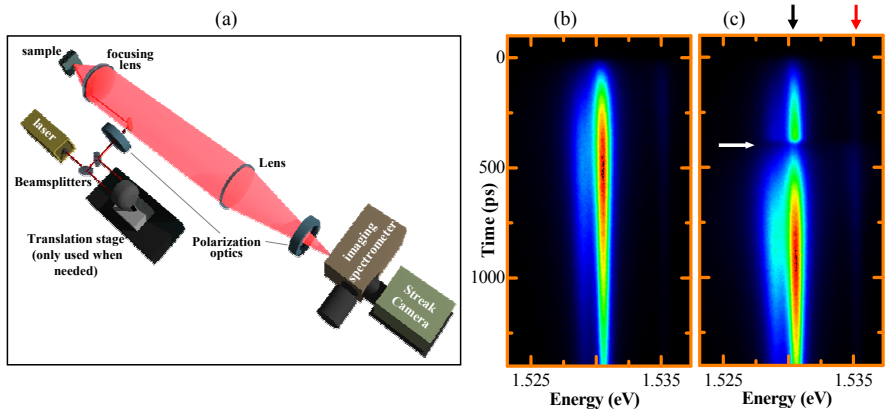}
\caption{(a) Schematics of the excitation and detection experimental set-up. (b)-(c) Streak camera images of the single QW \emph{hh-}PL under one pulse
excitation (b) and under two consecutive pulses excitation, $P_{I}$ and $P_{II}$,
with a delay between them of 400~ps (c), taken at a lattice temperature of 9~K. The color scales
are normalized in each panel. The white arrow indicates the arrival
of $P_{II}$ and the subsequent formation of a dip in the \emph{hh-}exciton
PL emission. The black (red) arrow mark the emission from \emph{hh}
(\emph{lh})-excitons.}
\label{figSetupPL2pulses}
\end{figure}

The studies have been carried out on two samples based on GaAs/AlAs: the first one
was a heterostructure with a single wide QW (20 nm) grown at \emph{Laboratoire de Photonique et de Nanostructures} in France, the second one contains multiple
(50) narrow QWs (7.7 nm), grown at the \emph{Paul Drude Institute} in Germany. Similar results were also found in a multiple InGaAs/GaAs QW
sample (10 nm), grown at the \emph{University of Sheffield}, evidencing the generality of the phenomena presented in this section.
In all cases the samples were kept in a cold finger cryostat at 9~K. We employed the time resolved
PL excitation and detection configurations in a back reflection geometry depicted in figure~\ref{figSetupPL2pulses}~(a). Excitation was performed with a Ti:Al$_{2}$O$_{3}$ laser that produced 1.5~ps long pulses with energy 26~meV above the heavy-hole (\emph{hh}) exciton. We use two consecutive pulses ($P_{I}$ and
$P_{II}$), whose power and delay can be independently controlled by means of attenuators and a delay unit. Both pulses arrive at the sample at the same excitation spot ($\sim$20~$\mu$m in diameter).
The PL from the excitons and the electron-hole recombination is collected by a streak camera attached to a spectrometer,with an overall time and energy resolution of 15 ps and 0.2 meV, respectively

Figure \ref{figSetupPL2pulses}~(b) and (c) show streak camera images of the
excitonic emission from the single QW sample after excitation by one single
pulse of 70~$\mu$W and by two identical consecutive
pulses delayed by 400 ps (70~$\mu$W each), respectively. In the latter
case, a clear dip appears in the emission of the \emph{hh} exciton
(black arrow) at the time of arrival of $P_{II}$ (white
arrow).

%***********************************************************************************
%***********************************************************************************

\begin{figure}
\centering\includegraphics[clip,width=0.9\textwidth]{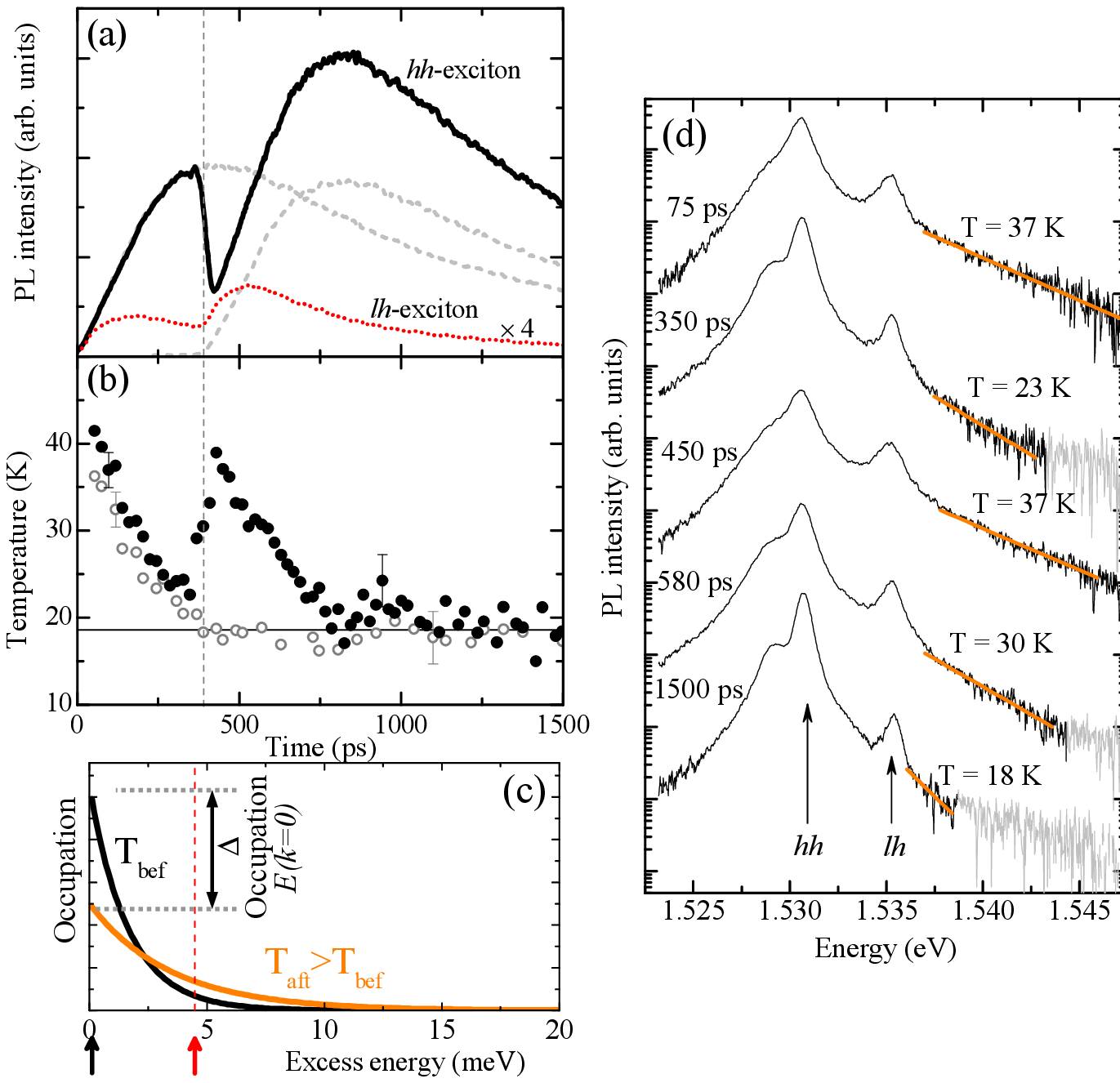}
\caption{(a) Photoluminescence dynamics of the single QW \emph{hh-}exciton
under one pulse excitation (grey dashed lines) and under two consecutive
pulses excitation (black solid line; conditions of figure~\ref{figSetupPL2pulses}; lattice temperature: 9~K).
In red dotted line the emission of the \emph{lh} excitons (enhanced
by a factor of 4) is presented under two pulses excitation. (b) Measured
electron-hole temperature for a single (open points) and double (solid
points) pulse experiment. (c) Exciton occupation right before (black
line) and right after (orange line) the arrival of $P_{II}$; the
black (red) arrow indicates the energy of the $k=0$ heavy (light)-hole
exciton. (d) Spectra taken at different delays (curves have been rigidly offset), showing the \emph{hh-} and \emph{lh-}exciton emission. The low energy shoulder to the \emph{hh-}exciton line corresponds to negative trions formed from residual electrons present in the sample. The
thick solid lines are fits to the free electron-hole pair recombination
following a Maxwell-Boltzmann distribution, from which the temperature
of the carriers can be extracted. The second pulse reaches
the sample at a delay of 400~ps~\cite{Amo2008b}.}

\label{Fig2pulses-1}
\end{figure}

Figure~\ref{Fig2pulses-1}~(a) shows in detail the time evolution of
the \emph{hh} exciton PL when the two pulses excite the QW independently
(dashed lines) and when both excite it jointly (solid line). The emission
from the light-hole (\emph{lh}) exciton can also be detected 4.7~meV
above the \emph{hh} exciton (red dotted line; red arrow in figure~\ref{figSetupPL2pulses}).
In contrast to the \emph{hh}, the \emph{lh} dynamics do not show the
appearance of a dip in the PL at the time of arrival of $P_{II}$.

%***********************************************************************************
%***********************************************************************************

\subsection{Origin of the dip: thermodynamical model\label{sub:Origin-of-the-dip}}

\begin{figure}
\centering\includegraphics[clip,width=1\textwidth]{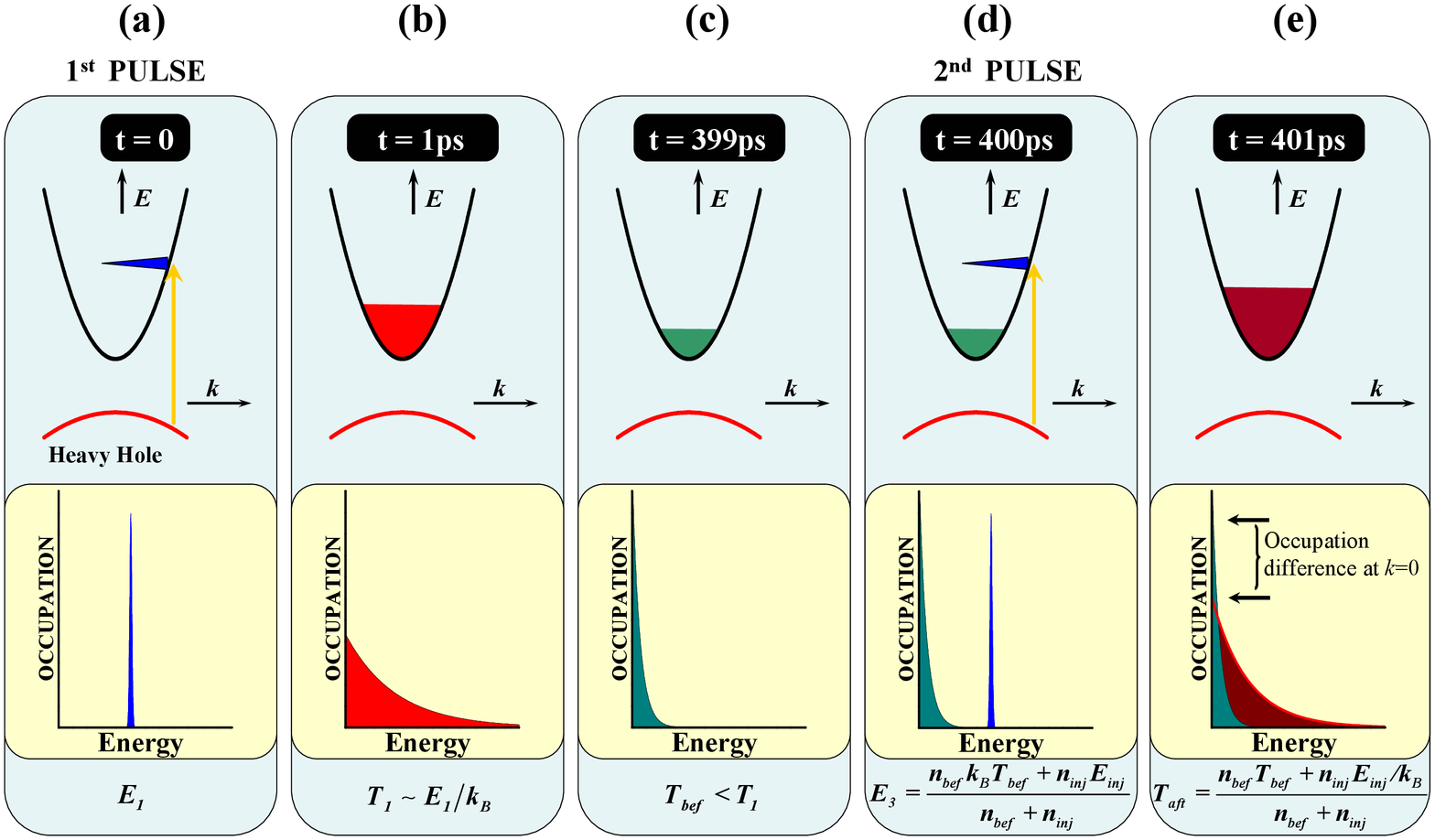}
\caption{Schematic time evolution of the electron distributions in a two pulses
experiment (holes are not shown for simplicity). See text for details.}

\label{Fig2PulsesThermal}
\end{figure}

The origin of the dip can be explained considering the redistribution
of carriers in the bands that takes place after the absorption of
$P_{II}$. To account quantitatively for the observed magnitude of
the dip depth, we have developed a dynamical quasi-equilibrium thermodynamical
model of the carriers in the QW \cite{Amo2008b}, which is sketched in figure~\ref{Fig2PulsesThermal}. The absorption of the first non-resonant pulse creates a non-thermal
population of electrons and holes at high energies in the conduction
and valence bands {[}figure~\ref{Fig2PulsesThermal}~(a)]. In
a time scale of the order of 200~fs the carriers distribute in the
bands achieving a well defined temperature, higher than the lattice
temperature {[}panel~(b)]~\cite{Knox1992,Rota1993b}. The carrier
densities considered in this section are far from the degeneration
limit, and the electron population can be well described by a Maxwell-Boltzmann
distribution function:
%-----------------------------------------------
\begin{equation}
\centering f_{MB}(E)=\frac{n}{k_{B}T}\frac{1}{DOS}\exp(-E/k_{B}T),
\label{eq:2pulsesBoltzman}\end{equation}
%-----------------------------------------------
where $k_{B}$ is Boltzmann's constant, $DOS$ is the carrier's density of states,
$E$ is the energy of the electrons above the bottom of the
conduction band, and $n$ and $T$ are their density and temperature, respectively. A similar description
can also be made for holes. As time evolves, the carrier distributions
cool down due to the interaction with phonons, and the populations
decrease due to pair recombination and exciton formation {[}panel~(c)]~\cite{Leo1988b}.
When $P_{II}$ reaches the QW new hot carriers are photoinjected ($n_{inj}$)
at high energy in the bands {[}$E_{inj}$, panel~(d)]. Due to efficient carrier-carrier
scattering, the newly created carriers and the pre-existing populations
rethermalize in a timescale given by the pulse duration, to a temperature
($T_{aft}$) higher than the carrier temperature right before the
absorption of $P_{II}$ ($T_{bef}$) {[}panel~(e)]. Thus, the effect
of the delayed pulse is to warm-up the carriers and increase their
concentration, in a time-scale shorter than a few picoseconds ($\sim$2~ps).

%***********************************************************************************
%***********************************************************************************

Simultaneously to the thermalization and cooling processes we have just described for the free carriers, electrons and holes bind to form excitons in a time-scale of the order of several hundreds of picoseconds, for the excitation densities and lattice temperature of our experiments~\cite{Szczytko2004,Bajoni2006b}. As discussed above, the exciton population obeys
also a Maxwell-Boltzmann distribution law. Recent studies have successfully developed kinetic theories, fitted to experimental data, based on the assumption that excitons and the
coexisting electron-hole plasma have the same temperature at all times \cite{Szczytko2004,Chatterjee2004,Szczytko2005,Hoyer2005}. 
Bajoni \emph{et al.} \cite{Bajoni2006b} have actually measured independently
the temperature of the excitons and that of the electron-hole plasma,
in a sample similar to the single, wide QW structure used here.
They observe that both species do show the same temperature for times
larger than 200~ps after a pulsed non-resonant excitation. For shorter
times their temperatures differ and Bajoni and coworkers propose that the exciton populations have
not reached a thermal distribution, while the electon-hole plasma has thermalised in less than one picosecond. This argument agrees with the theoretical
predictions of Selbmann \emph{et al.} in~\cite{Selbmann1996}, where they discuss
that the exciton formation is favored for electrons lying with kinetic
energies around the LO-phonon, as LO-phonon assisted electron-hole
binding is the most favorable exciton formation mechanism. This fact, added to the slow exciton momentum
relaxation, would result in the non-thermalicity of the exciton population
in the first 200~ps. During this initial
time excitons are simultaneously forming, relaxing and thermalizing.

The situation in our two-pulses configuration is slightly different.
In our case $P_{II}$ reaches the sample more than 200~ps after the
arrival of $P_{I}$, and we can expect thermal equilibrium between
excitons and plasma at that time. Therefore, $P_{II}$ injects hot
electron-hole pairs in a system populated by already thermalized excitons.
In the first picoseconds, after the arrival of $P_{II}$, the rethermalized
electron-hole plasma just warms the preexisting excitons. However,
in contrast to the plasma population, due to the slow formation dynamics
of excitons ---two orders of magnitude slower than the excitation
pulse---, the exciton density is hardly altered within the pulse duration.
Thus, no additional non-thermalized excitons should be considered
during the time of arrival of $P_{II}$.

Following this argument, we can assume that excitons and the electron-hole plasma have the
same temperature at the time of arrival of $P_{II}$, and probably
for some time after that. Therefore, an analogous dynamics to that
of the free carriers takes place also for the exciton population:
during the arrival of $P_{II}$, the abrupt warming of the carriers
results in an ultrafast warming of the exciton population.

With these assumptions and using equation~(\ref{eq:2pulsesBoltzman}) the
appearance of the dip in the \emph{hh-}exciton PL can be understood.
The \emph{hh-}exciton occupation of the \emph{zero} momentum states
($k=0$) before/after the arrival of $P_{II}$ is given by:
%---------------------------------------------------------------------
\begin{equation}
\centering f_{MB}^{bef/aft}(0)=\frac{n'}{k_{B}T_{bef/aft}}\frac{1}{DOS_{exc}},\label{eq:2pulsesOccK0}\end{equation}
%---------------------------------------------------------------------
where $n'$ is the pre-existing density of excitons at the arrival
time of the pulse, $DOS_{exc}$ is the excitonic density of states, and $T_{bef/aft}$ is the exciton 
temperature (equal to that of carriers) before/after the arrival of $P_{II}$.

The PL intensity of the \emph{hh} excitons is directly proportional
to the occupation of states with $k$ close to \emph{zero}, as these
are the excitonic states that can couple to light. Therefore, modifications
in the occupation of these states at the arrival of the delayed pulse
{[}see figure~\ref{Fig2pulses-1}~(c)] induce changes in the PL. In
particular, the abrupt warming of the carriers produced by the arrival
of $P_{II}$ ($T_{aft}>T_{bef}$) results in an abrupt drop of the
$k=0$ populations ($f_{MB}^{aft}(0)<f_{MB}^{bef}(0)$) and consequently
in an ultrafast quenching of the \emph{hh}-PL, as born out by our
experiments (see figure~\ref{Fig2pulses-1}).

The details of the exciton redistributions after the arrival of $P_{II}$
also accounts for the absence of a dip in the \emph{lh} emission.
The \emph{lh-}exciton energy is higher than that of the \emph{hh}
exciton and the increase in temperature of the excitons results in
a negligible change of the excitonic occupation at the $k=0$ \emph{lh}
states {[}as depicted in figure~\ref{Fig2pulses-1}~(c) for an energy
marked with the red arrow].

We can gain insight into the quantitative validity of
this quasi-equilibrium thermodynamical model, by directly measuring
the carrier temperature in the single, wide QW. Figure~\ref{Fig2pulses-1}~(d)
shows PL spectra detected at different times after the arrival of
$P_{I}$ ($P_{II}$
arrives at a delay of 400 ps). The spectra show the \emph{hh-} and \emph{lh-}exciton
 lines as well as the direct free electron-hole pair recombination
above the \emph{lh-}exciton energy. A Maxwellian fit can be performed
at the high energy tail (indicated by solid lines) to extract the
carrier temperature. Figure~\ref{Fig2pulses-1}~(b) shows the temperatures
of the electron-hole plasma for the conditions of figure~\ref{Fig2pulses-1}~(a).
The solid (open) dots corresponds to the double (single) pulse experiments.
In the case of the single pulse excitation, a monotonous cooling of
the plasma is observed, with a final equilibrium temperature ($\sim$19~K)
higher than the lattice temperature ($\sim$9~K)~\cite{Leo1988b,Yoon1996}.
In the two-pulses experiment, an abrupt warming of the plasma, and
consequently of the excitons, can be observed at the time of arrival
of $P_{II}$. At longer times ($>1$~ns) the same temperature as that
obtained in a single pulse experiment is reached.

The measured temperature of the carriers enables us to predict the
dip depth following our model. A relative dip depth can be defined
as $r\equiv(I_{bef}-I_{aft})/I_{bef}$, where $I_{bef}$ ($I_{aft}$)
is the intensity of the \emph{hh-}exciton PL right before (after)
the arrival of $P_{II}$. From this definition and equation~(\ref{eq:2pulsesOccK0}),
the relative dip depth can be related to the excitonic temperature
before and after the arrival of $P_{II}$ by:

%------------------------------------------------------------------------------------------------
\begin{equation}
r=\frac{f_{MB}^{bef}(0)-f_{MB}^{aft}(0)}{f_{MB}^{bef}(0)}=1-\frac{T_{bef}}{T_{aft}}.\label{eq:2pulsesRvsT}\end{equation}
%------------------------------------------------------------------------------------------------

\begin{figure}
\centering\includegraphics[clip,width=0.4\textwidth]{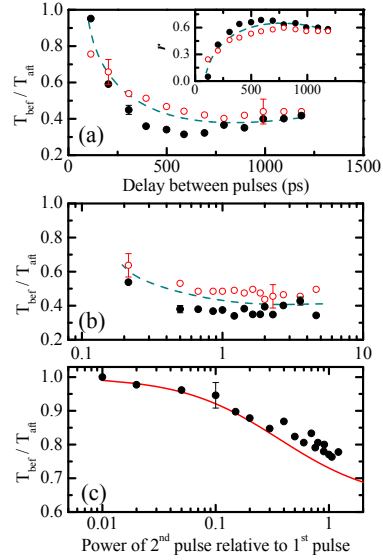}
\caption{Temperature ratio (just before the second pulse/just after the pulse)
in the single GaAs/AlAs QW sample as a function of (a) \emph{delay}
between pulses for a fixed power of both pulses (70~$\mu$W), and
(b) \emph{power} of $P_{II}$ for a fixed delay between pulses of
400 ps and fixed power of $P_{I}$ (70~$\mu$W). The solid (open)
dots quantify the temperature ratio as directly measured from the
PL (predicted from the model using the value of the measured dip depth). The dashed lines are guides to the
eye. The inset of (a) shows the corresponding relative dip depths
(\emph{r}). (c) Same as (b) for the multiple narrow QW sample with
delay between pulses of 300~ps; the solid line computes the model \cite{Amo2008b}.}

\label{Fig2pulses-3}
\end{figure}

The inset of figure~\ref{Fig2pulses-3}~(a) shows in solid dots ($\bullet$)
the values of $r$ directly measured from the \emph{hh-}exciton PL
as a function of the delay between the two pulses for the single QW.
The high values of $r$ demonstrate the capability of $P_{II}$ to
quench the PL. The open dots (\textcolor{red}{$\circ$}) depict $r$ obtained from
the measured carrier temperature ratio $T_{bef}/T_{aft}$ and equation~(\ref{eq:2pulsesRvsT}).
In a symmetrical manner, figure~\ref{Fig2pulses-3}~(a) depicts in
open dots $T_{bef}/T_{aft}$ as directly measured, while the solid
points compile the temperature ratio as obtained from the observed dip depth and equation~(\ref{eq:2pulsesRvsT}).
Figure~\ref{Fig2pulses-3}~(b) shows $T_{bef}/T_{aft}$ as a function
of the power of $P_{II}$ relative to that of $P_{I}$. In this case,
the signal to noise ratio of the spectra limited the lowest power
of $P_{II}$ for which the temperature of the electron-hole plasma
could be extracted. The good agreement between the temperature ratios
obtained from the relative dip depth in the \emph{hh} PL ($\bullet$)
and those measured directly (\textcolor{red}{$\circ$}) confirms the validity of our
model.

%***********************************************************************************
%***********************************************************************************

This is further demonstrated in figure~\ref{Fig2pulses-3}~(c) that
depicts $T_{bef}/T_{aft}$ obtained from the \emph{hh-}exciton PL for the
GaAs/AlAs narrow multiple QW sample as a function of the power of
$P_{II}$ (delay~= 300~ps) in the low power regime (solid points).
In this case the broader excitonic linewidth hinders the possibility
of extracting the carrier temperature from the spectra. Nonetheless,
our model reproduces quantitatively the observed experimental dependence
without the need of any adjustable parameters, as shown by the red
line, which plots $T_{bef}/T_{aft}$ obtained in the following way:
the temperature after $P_{II}$ is given by $T_{aft}=(n_{bef}T_{bef}+n_{inj}T^{*})/(n_{bef}+n_{inj})$,
where $n_{bef}$ is the density of carriers at the time of arrival
of $P_{II}$, determined from the power of $P_{I}$ and the PL decay-time;
$n_{inj}$ is the density photoinjected by $P_{II}$; $T^{*}$ and
$T_{bef}$, taken as 38~K and 24~K, respectively, are the initial
carrier temperature and that measured at a delay of 300~ps in the
single QW experiments (see figure~\ref{Fig2pulses-1}), which were
performed under very similar conditions to those in the multiple-QW
structure.

The experiments and model presented in this section demonstrate that ultrafast optical control
of the off-transients in QWs is possible. Additionally, although the results presented here concentrate
on the light emission from this kind of nanostructure, the physics and observed phenomena can
be directly extrapolated to more complicated systems such as semiconductor microcavities,
vertical cavity surface emitting lasers (VCSELs) or structures with active media of higher dimensionality,
like bulk direct-gap semiconductors. In particular, the relationship between
the dip depth and the carrier warming can be exploited to study the carrier-relaxation dynamics
in microcavities as we will show in section~\ref{polaritonslinear}.

%% file: polaritonsLinear.tex
\section{Relaxation of microcavity polaritons \label{polaritonslinear}}

Very interesting particles intimately related to the QW excitons are microcavity polaritons. These objects arise from the strong coupling between an exciton resonance and a photon mode confined in a two dimensional optical cavity. Morphologically, the optical cavity is conformed by a two-dimensional high finesse Fabry-Perot resonator composed of an upper and a lower Bragg reflector separated by a cavity spacer with a thickness $N/2$ times the wavelength $\lambda_c$ of the optical mode to be confined, with $N$ being an integer number. The Bragg reflectors are composed of alternating layers of two semiconductor materials with different refractive index with an optical length of $\lambda_c/4$. Typically, when the number of layers in each reflector is on the order of 25, reflectivities as high as 99.95\% can be achieved in GaAs based systems, giving rise to a confined photonic mode with a spectral width on the order of 0.1~meV.

%**********************************************************************************
\begin{figure}
\centering\includegraphics[width=0.95\textwidth]{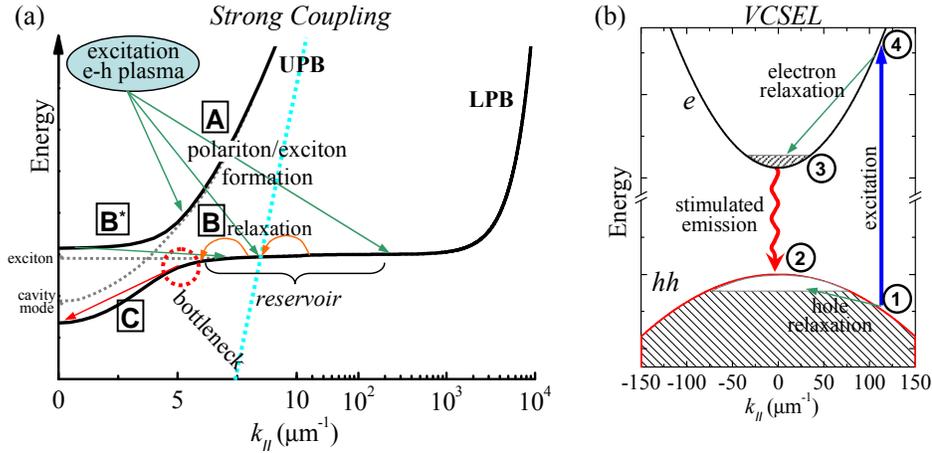}
\caption{(a) Polariton dispersion in the strong coupling regime (solid lines), and exciton and photon modes at slightly negative detuning in the absence of coupling (dashed lines). The carrier relaxation channels are depicted:
{[}A] polariton/exciton formation from the photocreated plasma of electrons
and holes, {[}B] polariton relaxation and thermalization within the
\emph{reservoir} states, {[}$\textrm{B}^{*}$] polariton relaxation
from the bottom of the UPB to the \emph{reservoir} states, {[}C] polariton
relaxation from the bottleneck to the bottom of the LPB. The dashed blue
line indicates the light cone. (b) Carrier relaxation and light emission in the weak-coupling
regime above the threshold for photon lasing (VCSEL regime). The numbers
indicate the leveling of the 4-level lasing system.}

\label{FigRelaxLPB}
\end{figure}
%**********************************************************************************

In the cavity spacer, where the confined electromagnetic field has its maxima, a single or a group of identical QWs is grown. The thickness of the spacer is designed so that the energy of the confined photon mode is equal or close to that of the QW exciton. Under these conditions, strong coupling between the photon and exciton modes can be achieved~\cite{Weisbuch1992}, giving rise to polaritons, the new eigenstates of the system which are a quantum mixture of exciton and photons. One of their most interesting characteristics is their rich dispersion, composed of two branches, the upper and lower polariton branch (UPB and LPB, respectively), which are shown in figure~\ref{FigRelaxLPB}~(a).

\subsection{Polariton condensates}

Due to their partially photonic character, polaritons in the momentum trap around $k=0$ have a very small mass, on the order of $10^{-5	}$ times the free electron mass. This characteristic added to the fact that polaritons are composite bosons have converted microcavities in an ideal system for the study of bosonic quantum degeneracy at temperatures ranging from a few kelvins to room temperature~\cite{Kavokin2007}. Among other observations, polaritons have evidenced lasing without inversion~\cite{Imamoglu1996,Dang1998,Christopoulos2007,Christmann2008,Bajoni2008}, Bose-Einstein condensation~\cite{Kasprzak2006,Balili2007}, long range order \cite{Kasprzak2006,Lai2007b}, appearance of quantized vortices~\cite{Lagoudakis2008,Sanvitto2009} and superfluidity~\cite{Amo2009,Amo2008}, as well as phenomena like optical parametric oscillation (OPO)~\cite{Stevenson2000,Romanelli2007} and amplification~\cite{Savvidis2000,Sanvitto2005b}, optical bistability~\cite{Baas2004,Demenev2008} and squeezing~\cite{Karr2004}. Many of these behaviours show particularities absent in other bosonic systems (like ultracold atomic condensates) due to the short polariton lifetime ($<$~10~ps). For instance, the steady state of the system is given by the interplay between pumping, relaxation and decay, and not by a thermodynamic equilibrium. Additionally, very rich spin physics are found in these systems, which can be directly accessed via the polarization control of the excitation and detected fields~\cite{Martin2007b,Shelykh2010}.

Several terms like \emph{condensate}, \emph{polariton laser} and \emph{Bose-Einstein condensate}, are used in the literature to refer to related 
but subtly different phenomena with respect to the coherent phases of polaritons in semiconductor microcavities. For the sake of clarity and to avoid ambiguities in the present manuscript, we would like to briefly address the meaning of these three concepts.

\emph{Condensate} is the most general of the three, and it is employed to designate a macroscopic ensemble of particles occupying the same quantum state. In this sense, no matter how we excite the system (on-resonance or out-of-resonance), a gas of polaritons in a state which has an occupation above 1, is a condensate~\cite{Keeling2007}.

\emph{Bose-Einstein condensation} makes reference to the creation of a condensate which is in thermodynamic equilibrium, i. e. with a well defined temperature and chemical potential. BEC cannot take place in a purely bi-dimensional system at finite temperatures. For this reason this term may only be employed for polaritons under very special circumstances, in particular when condensation takes place under non-resonant excitation, at positive detunings (to allow for the polariton gas to thermalize, as we will discuss below)~\cite{Deng2006,Kasprzak2008b,Marchetti2008}, and in the presence of an additional confinement~\cite{Malpuech2003,Keeling2004b,Kasprzak2006,Balili2007,Malpuech2007}. In a purely two-dimensional system, a Berezinski-Kosterlitz-Thouless (BKT) transition can take place at finite temperatures, with subtle differences with respect to BEC (for instance, in the decay of the spatial coherence with distance)~\cite{Pitaevskii2003}.

\emph{Polariton lasing} is employed when the polariton condensate is in a non-thermal state, that is, if a state with occupation above 1 is attained but the total ensemble of polaritons does not follow a thermal distribution~\cite{Malpuech2002,Szymanska2006,Christmann2008,Bajoni2008}. Let us note that though in polariton lasers the system is out of thermal equilibrium, much of the physics expected in a system in equilibrium can still be observed due to the extension of the coherence time, which becomes longer than the typical interaction times. This extension arises from the stimulated relaxation processes from \emph{reservoir} states to the macroscopically occupied condensed state~\cite{Love2008}. Finally, a related term is \emph{photon lasing}, which is observed in microcavities if coherent emission under out-of-resonance excitation takes place in the weak coupling regime, as in a VCSEL (see discussion below)~\cite{Butte2002a,Bloch2002,Szymanska2002,Bajoni2007,Bilykh2009}.

The experiments showing so far polariton condensation under out-of-resonance excitation can be ascribed to either BEC or polariton lasing depending if the gas can be described by a thermalized distribution or not, while the BKT transition has not yet been clearly demonstrated experimentally. In planar microcavities, polariton BEC has been claimed either in artificial~\cite{Balili2007} or natural traps formed during the growth process~\cite{Kasprzak2006,Sanvitto2009b,Krizhanovskii2009}, breaking the translational symmetry in the plane of the cavity and allowing for the spontaneous condensation at finite temperatures. However, it must be noted that the experimental determination of the thermal character of the gas is not always unambiguous~\cite{Bajoni2007,Nelsen2009}.

In the following sections we will mainly refer to the general concept of condensate, particularly in section~\ref{fluiddynamics}, where we describe experiments under resonant excitation. The high occupation of the addressed state is induced by the direct injection of polaritons via the excitation laser, and the gas does not present a thermal character.

\subsection{The bottleneck effect}

The most extended configuration employed to explore the phenomenon of spontaneous polariton condensation
is the use of the energy trap in momentum space present in the lower
polariton branch around $k=0$ under non-resonant excitation. If a sufficient amount of polaritons
accumulate in the quantum states at the bottom of the trap, condensation takes place.
%, which manifests in the spontaneous appearance of long range order, spectral narrowing, non-linear intensity emission \cite{Kasprzak2006} and an extended temporal coherence \cite{Krizhanovskii, delValle}
In order to wash out the coherence of the excitation source (a laser) and to be able to observe the spontaneous formation of a condensed phase, non-resonant optical excitation is required, expecting that relaxation
in the bands will destroy any trace of the initial coherence of the
injected carriers. As depicted in figure~\ref{FigRelaxLPB}~(a)
the non-resonantly created polaritons relax energy and momentum until
they reach the momentum-bottleneck region.

Relaxation from the bottleneck region to the bottom of the polariton
trap is a slow process at low excitation densities, as it requires
the simultaneous relaxation of significant amounts of energy and momentum.
LO-phonon assisted scattering cannot participate due to
energy conservation constraints, and only polariton-polariton~\cite{Tartakovskii2000,Senellart2000} and
acoustic phonon-polariton interaction can mediate in the polariton
relaxation~\cite{Malpuech2003}. The presence of an electron gas can also enhance the relaxation process~\cite{Malpuech2002b,Tartakovskii2003,Lagoudakis2003,Perrin2005b,Bajoni2006}.

Apart from the relaxation bottleneck,
another important constraint for the accumulation of polaritons at
the bottom of the trap is their reduced lifetime. Due to their photonic content, their lifetime 
is of the order of 1-10~ps in standard GaAs and CdTe based systems.
Faster relaxation times are required to achieve occupations at $k=0$
above 1 to trigger the condensation. The usual strategy has been to
increase the excitation density until the occupation of the ground
state is high enough, as long as the strong-coupling regime is not
destroyed by the high carrier density (due to exciton ionization~\cite{Bloch2002}).

This situation is achieved at low temperatures (5~K) in CdTe based microcavities~\cite{Richard2005,Kasprzak2006} and in GaN based systems at room temperature~\cite{Christopoulos2007,Christmann2008,Baumberg2008}, where the exciton binding energy is very high (around 25~meV in the former case, and 40~meV in the latter one). In the case of GaAs based microcavities, particularly relevant due to their high crystalline quality, the reduced exciton binding energy (around 10~meV) has not allowed, until recently, the observation of polariton condensation under non-resonant excitation: screening from the non-resonant injected carriers ionizes the polaritons at densities lower than those required to produce the condensation in the bottom of the trap. When polaritons ionize, the strong coupling is lost and the microcavity starts to behave as a VCSEL, with laser emission from the cavity mode~\cite{Bajoni2007}. A strategy to circumvent these difficulties has been the creation of a confining real space trap where polaritons condense far from the excitation point~\cite{Balili2007}. Recently, an improved microcavity design with an enhanced Rabi splitting has allowed the observation of condensation at excitation densities below those driving the system into weak coupling~\cite{Wertz2009}.

In this section we will study the transition from the strong to the weak coupling regimes, and the onset of photon lasing, in GaAs based microcavities under non-resonant excitation. By making use of a two pulses configuration as that of section~\ref{twopulsesQW} and the model given by equation~(\ref{eq:2pulsesRvsT}) we present a technique which allows the controlled warming of \emph{reservoir} polaritons, simultaneously providing insights on the relaxation mechanisms from the upper to the lower polariton branches.

\subsection{Polariton formation and relaxation}

The studied sample is a $3\lambda/2$ GaAs microcavity with two stacks of three In$_{0.06}$Ga$_{0.94}$As QWs, with a Rabi splitting of 6~meV, which was kept at 5~K. The sample was designed and grown at the \emph{University of Sheffield}. The wedged shape of the cavity spacer allows for the study of different exciton-cavity detunings [$\delta=E_c-E_x$, where $E_{c(x)}$ is the cavity (exciton) energy] just by choosing different positions on the sample. Single pulse excitation (1.5~ps long, at a repetition rate of 82~MHz) was performed at an energy of 1.63~eV, above the first reflectivity minimum of the stop band, in the continuum of electron-hole states of the GaAs cavity spacer, except in section~\ref{sub:Polariton-relaxation-UPB}
where, in some of the presented experiments, the excitation is resonant with the UPB close to its minimum. The excitation spot on the sample is kept at around 100~$\mu$m in diameter. The PL was energy and time resolved in a configuration analogous to that depicted in figure~\ref{figSetupPL2pulses}. The emission from the $k\approx0$ states was selected by means of a pinhole in the far field.

%**********************************************************************************
\begin{figure}
\centering\includegraphics[clip,width=0.95\textwidth]{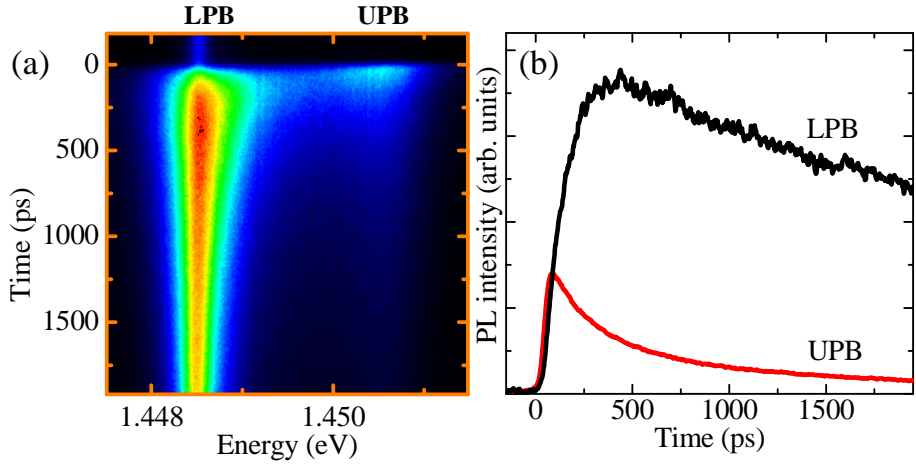}
\caption{(a) Streak camera image of the $k=0$ microcavity emission for a detuning
of $+7$~meV. (b) Time evolution traces of the upper and lower polariton
branches, extracted from (a). The UPB trace has been obtained by integration over a larger energy window than for the LPB trace, resulting in a reduced apparent noise.}

\label{FigStreakCameraPosDelta}
\end{figure}
%**********************************************************************************

Figure~\ref{FigStreakCameraPosDelta}~(a) shows a streak-camera image
of the $k=0$ emission at $\delta=+7$~meV after non-resonant excitation for low laser power (4~mW). Under these conditions
the LPB and UPB modes can be observed. In figure~\ref{FigStreakCameraPosDelta}~(b)
the time evolution traces of the upper and lower polaritons are shown.
The lower polariton branch presents slow dynamics, characterized by
a long rise and decay. In the case of the upper polaritons, the dynamics
are much faster.

The time evolution characteristics can be qualitatively understood
considering the phenomenological model described in figure~\ref{FigRelaxLPB}~(a).
Free electron-hole pairs are created by the non-resonant pulses in
the QWs. Analogously to the process described in section~\ref{sub:Origin-of-the-dip},
in less than a picosecond the electrons and holes achieve thermalized
distributions. Simultaneously, polaritons start their formation
process \{{[}A] in figure~\ref{FigRelaxLPB}~(a)\}, populating both
the UPB and the LPB. The cavity lifetime in the samples under study
is of the order of 2~ps. For states close to $k=0$, where polaritons have an important photonic component
($\sim50\%$) this implies that the polariton lifetime is on the order of 4~ps. For this reason, any polariton falling inside the dispersion
energy trap escapes very fast from the system. On the other hand,
polaritons above the bottleneck, outside the trap, possess a very high
excitonic component ($>90\%$) and a weak coupling to light. Additionally,
above a certain momentum, excitons lie outside the light cone {[}indicated
by a blue dahsed line in figure~\ref{FigRelaxLPB}~(a)] and do not couple to
light at all.%

Thus, the polariton lifetime is very different depending strongly
on the polariton momentum. Inside the trap the polariton modes are
strongly depleted, while above the bottleneck polaritons have long
lifetimes and dynamics similar to that of excitons, due to their high
excitonic component. Moreover, the bottleneck effect results in very
slow relaxations from the \emph{reservoir} to the bottom of the trap \{{[}C] in figure~\ref{FigRelaxLPB}(a)\}.
Given the very different lifetimes of polaritons inside and outside
the trap, and the bottleneck effect, it is straightforward to see
that any polariton relaxing from the \emph{reservoir} to the trap
immediately escapes from the system. The emission dynamics at
$k=0$, for non-resonant excitation, is mainly determined by the polariton
dynamics of the \emph{reservoir} states close to the bottleneck, as
these require the least amount of phonon or carrier scattering events
to relax their energy and momentum.

%Let us note that the trap region in momentum space is very small compared to the region of momenta over
%which thermalized polaritons extend above the bottleneck at the typical
%carrier temperatures of the photoluminescence experiments. An idea of the ratio between the number of carriers in each
%region can be obtained from the analysis of figure~\ref{FigNatMatKoch-2b},
%where exciton depletion within the light cone can be compared to the
%extension of the distribution of excitons in momentum space at temperatures
%similar to those in microcavity systems. 

The situation we have just described about the population distribution
in the LPB can be extended to any detuning, the only difference being
that at very positive detunings the LPB lifetime in the trap is increased,
and the bottleneck effect is reduced. Eventually, at very positive
detuning, the bare exciton dynamics are recovered. Still, states
close to $k=0$ are significantly depleted as compared to the states
outside the light cone~\cite{Koch2006}.

According to the arguments presented in the previous paragraphs, the
LPB $k=0$ polariton dynamics under non-resonant excitation reflects
the exciton dynamics at the \emph{reservoir} states. In this sense,
the long rise time depicted in figure~\ref{FigStreakCameraPosDelta}~(b)
is caused by the slow exciton formation, while the decay is mainly
characterized by the long exciton lifetime. This interpretation is
consistent with calculations of the polariton dynamics at different
temperatures and detunings~\cite{Savona1999}.

The UPB dynamics can also be explained attending to the sketch of
figure~\ref{FigRelaxLPB}~(a). The upper branch polaritons are very fast depleted due to either their
high photonic component (positive $\delta$) or to their strong interaction and fast relaxation to \emph{reservoir} lower branch polaritons
\{{[}$\textrm{B}^{*}$] in figure~\ref{FigRelaxLPB}~(a)\}~\cite{Sermage1996,Bloch1997}.
The large upper polariton linewidth is a manifestation of these interactions. 

The dynamics of the $k=0$ states of the UPB depicted in figure~\ref{FigStreakCameraPosDelta}~(b),
at a positive detuning on the order of the Rabi splitting ($\delta=+7$~meV), is a manifestation
of the two above mentioned fast depletion channels.
% The rise and decay
%UPB-dynamics depicted in figure~\ref{FigStreakCameraPosDelta}~(b),
%at a considerable positive detuning ($\delta=+7$~meV), are determined
%by two factors: (\emph{i}) at this positive detuning, the UPB is of
%highly photonic character, with a very short state lifetime; (\emph{ii})
%the relaxation channel {[}$\textrm{B}^{*}$] in figure~\ref{FigRelaxLPB}(a)
%communicates the \emph{reservoir} states with the UPB. In this way,
In particular, both the fast rise and the slow decay can be understood as 
a consequence of the quasi-thermal equilibrium between the \emph{reservoir}
states and the UPB: the fast rise results from the achievement
of a polariton distribution in the \emph{reservoir}, while the decay
reflects the cooling of that distribution and that of the total \emph{reservoir}
population itself.

\subsection{Transition from the strong to the weak coupling regimes}

\begin{figure}
\centering\includegraphics[width=1\textwidth]{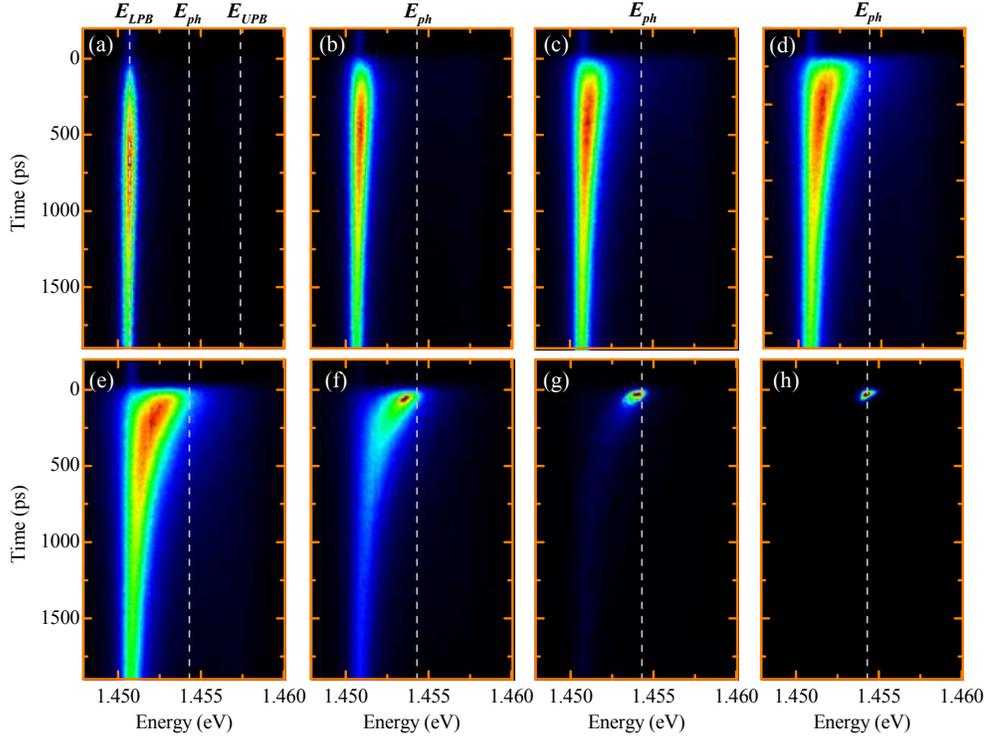}
\caption{Microcavity luminescence at $k=0$ at 5~K after pulsed non-resonant
excitation above the first minimum of the stop band at different powers:
(a)~1~mW, (b)~6~mW, (c)~10~mW, (d)~15~mW, (e)~20~mW, (f)~25~mW,
(g)~30~mW, (h)~45~mW. The dotted lines in (a) depict
the energy of the $k=0$ states of the LPB (1.4507~eV), cavity mode
(1.4542~eV) and UPB (1.4574~eV).}

\label{FigStreakCameraPowDepD0}
\end{figure}

The microcavity PL dynamics dramatically change when the excitation density is varied. This situation is explored in figure~\ref{FigStreakCameraPowDepD0}, where the emission dynamics at $k=0$ is shown as a function of excitation power for a spot of $\sim100$~$\mu$m in diameter, and at detuning very close to \emph{zero} ($\delta=+0.8$~meV). At very low power {[}figure~\ref{FigStreakCameraPowDepD0}~(a)], a narrow
emission ($\sim500$~$\mu$eV) is observed at the energy of the LPB.
Also emission from the UPB can be observed with the appropriate setting
of the \emph{z}-scale, but it is not shown in this figure. As the
carrier density is increased {[}figure~\ref{FigStreakCameraPowDepD0}~(b)-(c)], the polariton linewidth increases due
to the enhancement of the polariton-polariton and polariton-carrier
interactions~\cite{Ciuti1998,Baars2000,Huynh2003}. These interactions
are also responsible for the shortening of the rise time as they also
reduce the exciton formation time and ease the relaxation to the bottom
of the trap. In figures~\ref{FigStreakCameraPowDepD0}~(d)-(f) the transition
from the strong- to the weak-coupling is evidenced. At short times,
at these powers (and above), the high density of free carriers injected
by the non-resonant excitation pulse screens the electrons and holes
that form the excitons, leading to their ionization, and the system is driven
into the weak-coupling regime, characterized by emission at
the energy of the cavity mode ($E_{ph}$). At longer times, the number of carriers
in the system decreases due to recombination. In this way the screening
also decreases and the system goes back to the strong-coupling regime,
characterized by emission at the energy of the LPB and UPB, as evidenced
in figures~\ref{FigStreakCameraPowDepD0}~(d)-(g) at long times.

%***********************************************************
%***********************************************************

Once the strong coupling is lost, the dispersion relations of the
energy states of the system are given by those of the first electron
and heavy-hole subbands, as depicted in figure~\ref{FigRelaxLPB}~(b).
A four level system can be considered, as indicated in the figure,
and population inversion between levels 3 and 2 is easily achieved
as the electron and hole relaxation to the bottom/top of their bands
is very efficient due to carrier-carrier interaction~\cite{Rota1993b}.
In this case the Bragg mirrors act as very efficient resonators with
a photonic mode very close to the bandgap, and photon lasing is triggered,
with a very low power threshold~\cite{Pau1996,Butte2002a,Lagoudakis2004}.

Figures~\ref{FigStreakCameraPowDepD0}~(f)-(h) show how, as soon as
the system reaches the weak-coupling regime, photon stimulated emission
takes place and the microcavity is driven into VCSEL operation. Due
to the stimulated character of the recombination, the light emission
is very fast in this regime. Most available electron-hole pairs at
the cavity energy recombine in the first $\sim50$~ps.

This situation of photon lasing is very different to that of a  polariton BEC or a polariton
laser (\emph{plaser})~\cite{Imamoglu1996}. Optical emission in a \emph{plaser} or BEC
would come from the leakage out of the microcavity of the condensed polaritons at $k=0$.
Due to their bosonic nature, \emph{reservoir} polaritons with
large $k$ relax to the condensed state at the bottom of the LPB, via a process of final state stimulated scattering. The condensed
polaritons are in a well defined quantum state, and when they leak
out of the cavity they present very similar characteristics to those of
a conventional photon laser, i.e. monochromaticity, well defined phase,
directionality~\cite{Bajoni2007}. The fact that the stimulation and emission processes
are decoupled can lead, in principle, to \emph{plasing} devices without
excitation threshold~\cite{Imamoglu1996}.
%*****************************************************************************
\begin{figure}[t]
\centering\includegraphics[clip,width=0.5\textwidth]{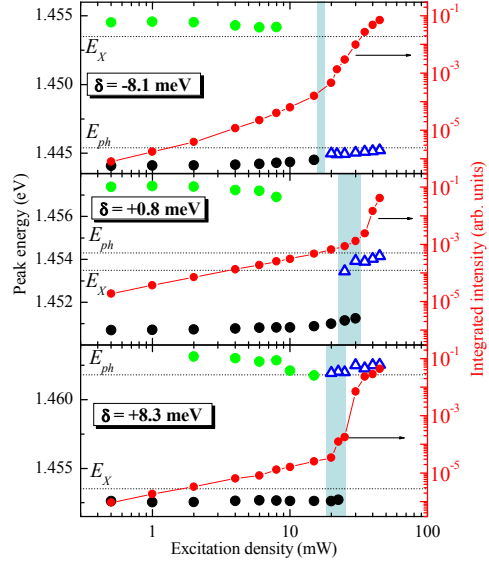}
\caption{Energy of the $k=0$ emission of the UPB (green dots), LPB (black
dots) and cavity mode (blue open triangles) ---left scales---, and
total integrated emission (red dots) ---right scales--- as a function
of non-resonant pulse power for several values of $\delta$. The
blue area shows the transition from the strong to the weak coupling
regimes. The dotted lines indicate the energy of the bare cavity ($E_{ph}$) and
exciton modes ($E_{X}$).}

\label{FigPowerDepDifDelta}
\end{figure}
%*****************************************************************************
Figure~\ref{FigStreakCameraPowDepD0}~(f) shows that, at short times
after excitation, emission from weakly (at the photon mode) and strongly
(at the LPB) coupled modes coexist. This phenomenon can be observed
in experiments at very different detunings. Figure~\ref{FigPowerDepDifDelta}
shows the emission energy (green and black dots) as well as the intensity
(red dots) dependence on pulsed-excitation power at several detunings
in the sample. At low excitation density, the microcavity is in the
strong-coupling regime with emission occurring at the LPB and UPB
energies, with a linear dependence of the photoluminescence intensity
on excitation density. The light blue areas depict the transition
from the strong to the weak coupling at high densities. Analogously
to the analysis carried on figure~\ref{FigStreakCameraPowDepD0}, the
transition threshold can be identified by the shift of the emission
to the cavity mode states (blue triangles). The appearance of emission
at the cavity mode is accompanied by the onset of a superlinear dependence
of the photoluminescence intensity on excitation power. Such a superlinear
behavior is characteristic of lasing systems right at the threshold
density under non-resonant excitation~\cite{Sargent1987,Doan2005}. At higher excitation intensities
the linear behavior is recovered, as hinted in the lower panel of figure~\ref{FigPowerDepDifDelta}.

The width of the light blue areas in figure~\ref{FigPowerDepDifDelta}
indicates the excitation power range over which strongly and weakly
coupled modes can be observed. At detunings close to \emph{zero},
where the strongly and weakly coupled modes are furthest apart in
energy the coexistence is clearly observed. In this case the coexistence
region could probably be extended to higher densities if a larger
dynamic range would have been available in these experiments, as the emission from the
photon lasing mode increases non-linearly while that of the coupled modes
increase linearly with density. Let us also mention that studies in
the same system evidence that the coexistence of strong and weak coupling
takes place on emission areas smaller than $\sim10$~$\mu$m in diametre~\cite{Ballarini2007}.

\subsection{Polariton relaxation from the UPB in the strong coupling regime\label{sub:Polariton-relaxation-UPB}}\sectionmark{Polariton relaxation from the UPB}

\begin{figure}
\includegraphics[clip,width=1\textwidth]{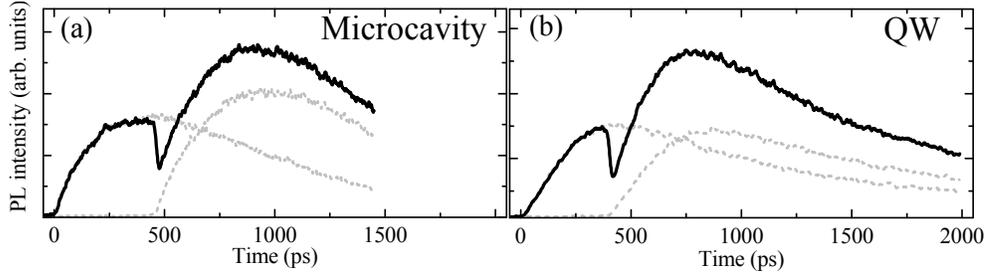}
\caption{(a) PL dynamics of the $k=0$ lower branch polaritons in the microcavity
under one pulse excitation (grey dashed lines) and under two consecutive
pulses excitation (black solid line; delay between pulses: 450~ps;
power of both pulses: 0.1~mW). (b) Same as (a) for the PL dynamics
of the bare QW excitons in an identical microcavity without top mirror.
In this case the delay between pulses is 400~ps and the power of
each pulse is 0.3~mW.}

\label{Fig2pulsesMicrocavity}
\end{figure}

The $k=0$ lower-branch polariton dynamics is mostly determined by
the dynamics of the \emph{reservoir} polaritons, as it was pointed out in
the discussion of the polariton relaxation channels {[}figure~\ref{FigRelaxLPB}~(a)]
and evidenced by the slow dynamics shown in figure~\ref{FigStreakCameraPowDepD0}~(a).
In this section we will gain insight into this relationship through
the results of a two pulses experiment analogous to those described
in section~\ref{twopulsesQW}. In
those experiments, two delayed excitation pulses $P_{I}$ and $P_{II}$
reach the sample on the same excitation spot, $P_{II}$ inducing an
ultrafast warming of the exciton population created by $P_{I}$. Figure~\ref{Fig2pulsesMicrocavity}~(a)
shows the LPB $k=0$ emission under such a configuration for a spot
on the sample at $\delta=0$. In this case both pulses excite the
sample non-resonantly (above the QW electron-hole continuum) with
the same power (0.1~mW) and a delay between them of 450~ps. Analogously
to the results presented in section~\ref{twopulsesQW},
a quench of the photoluminescence is observed at the arrival of the
second pulse. $P_{I}$ creates an electron-hole pair population that
decays into \emph{reservoir} excitons, which eventually relax to the
bottom of the polariton dispersion trap giving rise to the early emission
shown in figure~\ref{Fig2pulsesMicrocavity}~(a). The arrival of $P_{II}$
induces an ultrafast warming of the exciton distribution in the \emph{reservoir}
producing an abrupt decrease of the exciton populations at the bottleneck
and a subsequent quench of the $k=0$ luminescence.

%***************************************************************************
%***************************************************************************

In order to stress that the physics behind the observed dip are fully determined
by the redistribution of excitons in the \emph{reservoir}, we have also performed
a similar two pulses experiment in an identical microcavity sample
that has been processed (chemical etching) in order to remove the
top Bragg mirror. In this way, the luminescence from the bare QWs can
be accessed. Figure~\ref{Fig2pulsesMicrocavity}~(b) shows the QW
emission under the same non-resonant two pulses experiment (in this
case with a delay between pulses of 400~ps). The dynamics of the
PL under a single pulse excitation (grey dashed lines) are alike in
the microcavity and the QW, evidencing that they have the same origin.
For a similar delay between pulses and ratio of the powers of $P_{I}$
and $P_{II}$, the magnitude of the dip is also very similar in both
cases, which is what would be expected from the model described in
section~\ref{sub:Origin-of-the-dip} if the dip is caused by the redistribution
of polaritons/excitons.

\begin{figure}
\includegraphics[clip,width=1\textwidth]{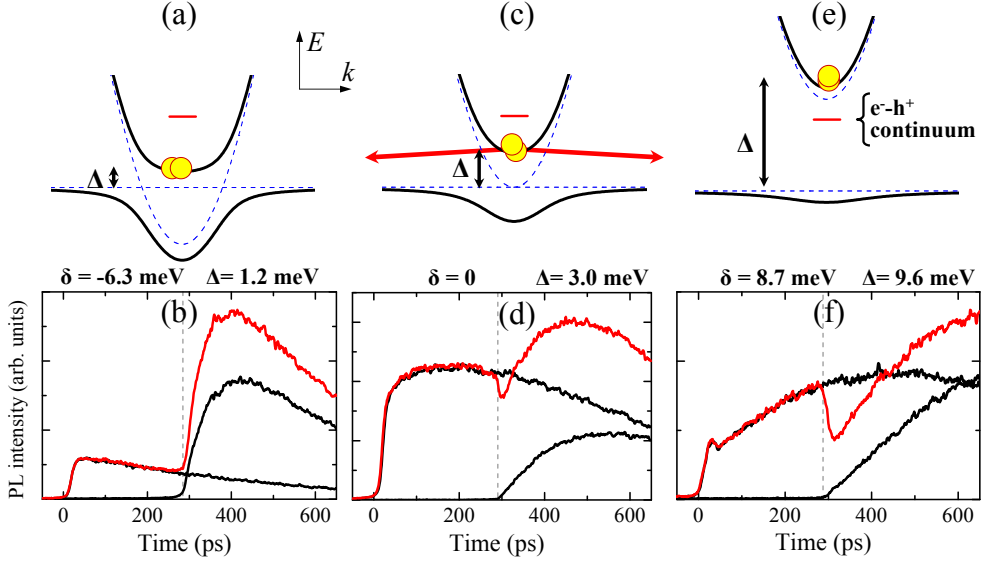}
\caption{Upper panels: polariton dispersion (solid lines) at $\delta=-6.3$~meV
(a), $\delta=0$ (b), $\delta=+8.7$~meV (c). The blue dashed lines
depict the dispersion of the uncoupled excitons and cavity photons.
The short red line indicates the energy of the electron-hole continuum
in the QWs. The yellow circles indicate the excitation energy and
momentum. The red arrows in (c) indicate the upper polariton scattering
into \emph{reservoir} states. Lower panels: PL emission at $k=0$
in the LPB under one pulse excitation (black lines) and under two
consecutive pulses excitation (red lines) at the detunings indicated
in the upper panels. In each case the power of both pulses is equal, and the delay between them is 285~ps.
The first pulse is linearly polarized, while the second one is $\sigma^{+}$
circularly polarized; only the $\sigma^{-}$ component of the emission
is detected (see text for details).}

\label{Fig2pulsesMicrocavityDiffDelta}
\end{figure}

Taking advantage of the model described in section~\ref{sub:Origin-of-the-dip}
and the relation between dip depth and relative temperature change
of the excitonic distribution [equation~(\ref{eq:2pulsesRvsT})], we are
going to study the relaxation of resonantly created upper-branch polaritons.
In this case the excitation will no longer be non-resonant above the
first minimum of the stop band. Here we will describe two pulses experiments
in which polaritons are resonantly injected in an UPB state close
to $k=0$. Figure~\ref{Fig2pulsesMicrocavityDiffDelta} shows the
$k=0$, LPB emission under excitation with two independent (black lines)
and two consecutive (red lines) pulses resonant with the aforementioned
UPB state, at different detunings. The delay between the pulses is 285~ps. The first pulse is linearly polarized,
the second one is $\sigma^{+}$~right-circularly polarized, and the $\sigma^{-}$~left-circular polarized emission
is detected. We use this configuration for reasons that will be explained
below. Control of the polarization of the excitation and detection
paths is performed by use of appropriate combinations of linear polarizers
and quarter waveplates.

%*********************************************************************************
%*********************************************************************************

Let us first concentrate on the dynamics when just one of the pulses, $P_{I}$, arrives to the sample (black lines in the lower panels of figure~\ref{Fig2pulsesMicrocavityDiffDelta}). In these experiments $P_{I}$, linearly polarized, resonantly injects
polaritons at the bottom of the UPB (see figure~\ref{Fig2pulsesMicrocavityDiffDelta}
upper panels). The direct relaxation via phonon emission to the $k\approx0$
LPB states results in the very fast rise of the PL observed right
after the arrival of $P_{I}$~\cite{Sermage1996}. Nonetheless,
a significant portion of polaritons scatter to the \emph{reservoir},
conforming a thermalized distribution. Some of
these polaritons relax, via emission of acoustic phonons, towards $k=0$, where they escape from
the cavity giving rise to the decay observed at later times~\cite{Sermage1996,Bloch1997}. When a second
delayed pulse reaches the sample, at the same energy and momentum
as $P_{I}$, the photocreated upper polaritons relax again to the
\emph{reservoir} states with an excess energy given by the difference
between the upper polariton energy and the \emph{reservoir} energy
(bare excitonic energy), which is indicated by $\Delta$ in figure~\ref{Fig2pulsesMicrocavityDiffDelta}.
The second pulse is $\sigma^{+}$ circularly polarized (injecting
spin-up polaritons), while only the $\sigma^{-}$ circularly polarized
emission is detected (from the escape of spin-down polaritons). In
this way the direct relaxation of the polaritons injected by $P_{II}$
from the UPB to the $k=0$, LPB states is disregarded, as this phonon
mediated process does not change the polariton spin. In contrast,
the \emph{reservoir} polaritons do flip their spin via usual exciton
spin flip mechanisms~\cite{Damen1991,Martin2002b}, and any change in their distribution
is reflected in the $\sigma^{-}$ PL emission at $k=0$. The slow
rise observed in the lower panels of figure~\ref{Fig2pulsesMicrocavityDiffDelta}
after the arrival of $P_{II}$ reflects the time required for \emph{reservoir}
polariton to flip their spin.

The red line in the lower panels of figure~\ref{Fig2pulsesMicrocavityDiffDelta} shows
the LPB $k=0$ emission when the two consecutive pulses reach the
sample. In panel~(d) a dip in
the PL can be observed indicating that the upper polaritons generated
by $P_{II}$ very efficiently relax towards the \emph{reservoir} states
{[}red arrows in (c)] changing significantly their distribution. By
measuring the relative magnitude of the dip depth, and making use
of equation~(\ref{eq:2pulsesRvsT}) we can calculate the temperature increase
of the \emph{reservoir} polaritons induced by the injection of upper branch
polaritons by $P_{II}$. In the case of $\delta=0$ {[}figure~\ref{Fig2pulsesMicrocavityDiffDelta}~(c)-(d)],
the temperature increase is given by $T_{aft}=1.2\times T_{bef}$.
For a larger detuning, as that depicted in figure~\ref{Fig2pulsesMicrocavityDiffDelta}~(e)-(f),
$\Delta$ is significantly increased, and the carriers injected by
$P_{II}$ are at higher temperature and warm stronger the exciton \emph{reservoir} ($T_{aft}=2.1\times T_{bef}$),
inducing a larger dip. In this case, additionally, the UPB lies within the electron-hole
continuum, and $P_{II}$ injects electron-hole pairs in the system,
which warm the \emph{reservoir} excitons even more efficiently than upper-branch
polaritons. Figure~\ref{Fig2pulsesMicrocavityDiffDelta}~(a)-(b) depicts
the case of negative detuning. In this case $\Delta$ is very
small, and the negligible excess energy of the polaritons injected
by $P_{II}$ results in a negligible warming of the \emph{reservoir}.

These results show that upper branch and \emph{reservoir} polaritons
form part of an ensemble in thermal quasiequilibrium. Even though there is a gap in momentum
between the two populations, they strongly interact, making possible the thermal control
of one of the two populations by changing the distribution of the other.
These properties and the well established relationship between dip depth and temperature change
could be an asset to investigate
phase transitions, such as condensation of indirect excitons~\cite{Butov2002} or polariton BECs~\cite{Kasprzak2006}, in
which the carrier temperature plays an important role.

%% file: fluiddynamics.tex
\section{Fluid dynamics of microcavity polaritons}
\label{fluiddynamics}

In the previous section we presented experiments under non-resonant
excitation in an InGaAs/GaAs/AlGaAs based microcavity. The emission
dynamics in the strong-coupling regime are well explained by free
carriers and polariton relaxation mechanisms, giving rise to
incoherent populations of polariton. Even though only the emission from the $k=0$
states has been so far considered, in this incoherent regime polaritons
distribute along the polariton dispersion giving rise to incoherent
luminescence from all states within the light cone.

As shown in the experimental results of section~\ref{polaritonslinear},
under highly non-resonant excitation (via photocreation of free electrons
and holes) in the GaAs based studied microcavity it is not possible to reach
a quantum condensed phase of polaritons~\cite{Bajoni2007}. In CdTe based microcavities
a Bose-Einstein condensed state has been observed by Kasprzak and
co-workers under such excitation conditions~\cite{Kasprzak2006}. In
 GaAs based microcavities condensation has been observed under
direct injection of \emph{reservoir} polaritons~\cite{Lai2007b,Deng2007}, within an externally created trap in real space~\cite{Balili2007} and, in a cavity designed with an enhanced Rabi splitting by the addition of numerous QWs to a high finesse structure~\cite{Wertz2009}.

The importance of these experiments is that a phase transition from
an incoherent state 
to a coherent condensed state of polaritons with \emph{zero} linear momentum is observed~\cite{Laussy2004b,Laussy2006,delValle2009}.
Very interesting properties distinctive of a condensed phase have
been reported for this \emph{zero} momentum state. In fact, experimental
observations of spectral and momentum narrowing, and
long-range order~\cite{Kasprzak2006,Balili2007,Lai2007b} have been used as proof of a Bose-Einstein type
phase transition in polaritons. However, despite these observations being clear
proof that polariton BECs can be formed in microcavities, they
do not differ significantly from what can be found in a pure photonic
laser~\cite{Bajoni2007}.

This subject has been recently addressed by several groups using different evidence. The formation of a condensed phase under non-resonant excitation should be evidenced by the observation of spontaneous symmetry breaking. In the case of polaritons in semiconductor microcavities it has been anticipated that this symmetry breaking should be manifested via the build-up of a well defined linear polarization of the emission of the condensate~\cite{Laussy2004b,Shelykh2006,Keeling2008b}, whose direction should be spontaneously chosen by the system~\cite{Read2009}. Indeed the formation of such linear polarization has been reported along with the above mentioned evidence of polariton condensation~\cite{Kasprzak2006,Balili2007,Lai2007b,Kasprzak2007}. However, in all these cases, the polarization appears pinned to some predefined crystalographic direction (see~\cite{Martin2005,Klopotowski2006}, and~\cite{Amo2009c} and references therein), which is a behaviour commonly found in VCSELs as well~\cite{Martin-Regalado1997}. Recently, in a bulk GaN based microcavity, where spin dependent exciton interactions are reduced, spontaneous selection of the polarization of the condensed phase was observed at room temperature~\cite{Baumberg2008}.

Spontaneous symmetry breaking in bosonic systems is often accompanied by
the appearance of a Goldstone mode in the spectrum of excitations. In
microcavity polaritons the Goldstone mode physics presents several
specificities, arising from the out-of-equilibrium character of polaritons, which can be accessed in the optical parametric
oscillation configuration~\cite{Wouters2007b}, when pumping the system
around the inflexion point of the dispersion. The symmetry breaking at the
onset of the OPO comes from the spontaneous selection of the phases of
signal and idler. While the sum of their phases must be equal to twice
that of the pump, their specific values are not fixed. Above the OPO
threshold, a spontaneous symmetry breaking sets the phase of signal and
idler, giving rise to the appearance of a Goldstone mode, as recently
reported experimentally~\cite{Ballarini2009}.

Another feature proper of a polariton condensates which is absent in a pure photonic laser is the phenomenology related to the interacting character of polaritons. In this sense, a detailed analysis of the particle number fluctuations~\cite{Love2008,Kasprzak2008} and the dynamics of coherence formation~\cite{delValle2009,Nardin2009}, provide proofs of the condensate/polariton lasing character of the system above some threshold.

An additional differential landmark between a Bose-Einstein condensate of interacting
polaritons and a condensate of non-interacting photons (a laser),
would be the observation of superfluid properties in the polariton
condensate~\cite{Kavokin2003c,Carusotto2004}. Superfluidity is a property emerging from inter-particle interactions in bosonic condensates, and gives rise to a varied phenomenology~\cite{Leggett2001,Keeling2009}. As introduced by Landau in the context of He-II liquids~\cite{Landau1986}, the two main characteristics from which almost all the phenomenology associated to superfluidity can be explained are: (\emph{i}) the irrotational character of the fluid (${\bf \nabla}\times{\bf v}=0$, where {\bf v} is the fluid velocity), and (\emph{ii}) its linear spectrum of excitations around \emph{zero} momentum. The first feature gives rise to the formation of vortices with quantized angular momentum, which have been extensively observed in He-II~\cite{Hess1967,Packard1972,Williams1974}, atomic Bose-Einstein condensates~\cite{Chevy2000,Madison2000,Abo-Shaeer2001}, and recently in polariton fluids~\cite{Lagoudakis2008,Sanvitto2009,Wouters2009,Lagoudakis2009}.

A linear spectrum of excitations with a discontinuity of the slope at \emph{zero} momentum is the key ingredient to understand the frictionless phenomena of superfluids~\cite{Leggett2001}. The slope of such spectrum sets a critical fluid velocity below which perturbations in the fluid caused by static external potentials are not possible: a gap opens for the excitation of the fluid, and frictionless movement is attained. Above this critical velocity, external static potentials give rise to excitations which present specific phenomenology due to the macroscopic quantum nature of the fluid. In this case, scattering takes place giving rise to a $\check{\textrm{C}}$erenkov
like pattern~\cite{Carusotto2006}.

Precise measurements of the critical velocity in superfluid He-II indirectly evidence the linear spectrum of excitations~\cite{Allum1977,Ellis1980}. In atomic BECs it has been directly measured~\cite{Steinhauer2002}, while in polaritons, dispersions presented with linear fittings resembling the Bogoliubov spectrum, have been recently reported~\cite{Utsunomiya2008}. The persistence of fluid currents is another phenomenon which emerges as a consequence of the frictionless flow. The canonical experiment consists on the stirring of a superfluid in a static cylindrical container which will continue rotating indefinitely with the same angular frequency even when the stirring force has been removed. Observations of this behaviour have been reported in He-II~\cite{Vinen1961,Whitmore1968}, atomic condensates~\cite{Ryu2007} and, recently, in polariton systems~\cite{Sanvitto2009}.

In this section we will study superfluidity in polariton condensates in the sense of flow without friction. As in our PL experiments we have not direct access to the spectrum of excitations of the system, we will concentrate on the scattering properties of a moving polariton fluid encountering a localized potential barrier. 
As mentioned before, the behaviour of a moving condensate interacting with a static obstacle
presents very specific scattering properties, which depend on the fluid velocity with respect to that of the obstacle: below some critical speed, scattering is suppressed, while above it, $\check{\textrm{C}}$erenkov scattering is observed~\cite{Carusotto2004,Ciuti2005} as reported, for instance, in atomic condensates~\cite{Raman1999,Onofrio2000,Fort2005,Carusotto2006}.

A polariton condensate with \emph{zero} momentum, as those studied in references~\cite{Kasprzak2006,Balili2007,Lai2007b}, is not suitable
to study these phenomena. A state with well defined
non-\emph{zero} momentum should be employed so that it can interact with static
obstacles in the sample. For this purpose it is more appropriate to consider the configuration in which polaritons are created by a laser field resonant with a non-\emph{zero} momentum state of the LPB. Recent experimental results have shown evidence of superfluid motion of polariton under such conditions~\cite{Amo2009,Amo2008}.

Here, we will experimentally explore this situation by employing the same technique as in reference~\cite{Amo2009}. We
will show the details of this configuration based on the triggered optical parametric oscillator (TOPO), which enables the creation of macroscopically occupied polariton states with a well defined and controlled non-\emph{zero}
momentum. Then we will discuss experimental results on the interaction of these
quantum fluids with obstacles of different sizes found in the sample.

\subsection{Making polaritons flow\label{sub:Making-polaritons-flow}}

The most straightforward way to make polaritons flow would be by the
resonant, pulsed photocreation of polaritons in a state of well defined momentum in
the LPB. A pulsed source is essential if the dynamics of the polariton fluid
is to be studied. However this configuration presents many difficulties when
it comes to the detection of the polariton movement. First of all,
polaritons only live a few picoseconds before escaping the cavity
($\sim4$~ps). The temporal dynamics of polaritons after
a resonant pulsed laser excitation can be resolved detecting photon
arrival times from the sample. However, even the best detectors do
not allow for resolution better than 2-4~ps, when the polariton population of the excited state has almost completely
disappeared. Additionally, stray light from the resonant laser excitation may hinder
the detection of their movement in the cavity, even a few picoseconds
after the arrival of the pulse.

Very few experiments in the literature address the issue of polariton
movement. Freixanet, Sermage and coworkers~\cite{Freixanet2000,Sermage2001}
presented an experimental configuration in which polaritons with
a well defined momentum state were created by resonant pulsed excitation
of the LPB at a given angle of incidence. In order to avoid the aforementioned
difficulties, they made use of a pinhole to block the stray light
from the laser. In this way they restricted the observation to polariton
states with different momentum and energy to those resonantly addressed
by the laser. The observed states do not conform polariton fluids
with macroscopic occupations, as they are populated by the incoherent
scattering from other polariton states (mainly from the state excited
by the laser).

A related experiment is that of Langbein
\emph{et al.}~\cite{Langbein2007}. In this case a well defined polariton
state is not excited, as the pulsed source simultaneously populates
a plethora of states with all possible in-plane momenta at a given
energy in the LPB. The stray light from the laser is eliminated by
use of a confocal setup with a mask of the excitation spot. In any
case, propagation of the polaritons in a ring departing
from the excitation spot can be observed, with a very fast decay ($\sim4$~ps).

In those experiments,
besides the difficulties related to the very fast decay of the polariton
populations, the addressed states present an incoherent
nature.

Here we present an experimental configuration that allows us to create
a polariton fluid in a well defined energy and momentum state
while being able to observe its spatial evolution without any of the
difficulties related to the decay of the populations and of the stray
light from the laser.

%***************************************************************************
\begin{figure}
\centering\includegraphics[width=1\textwidth]{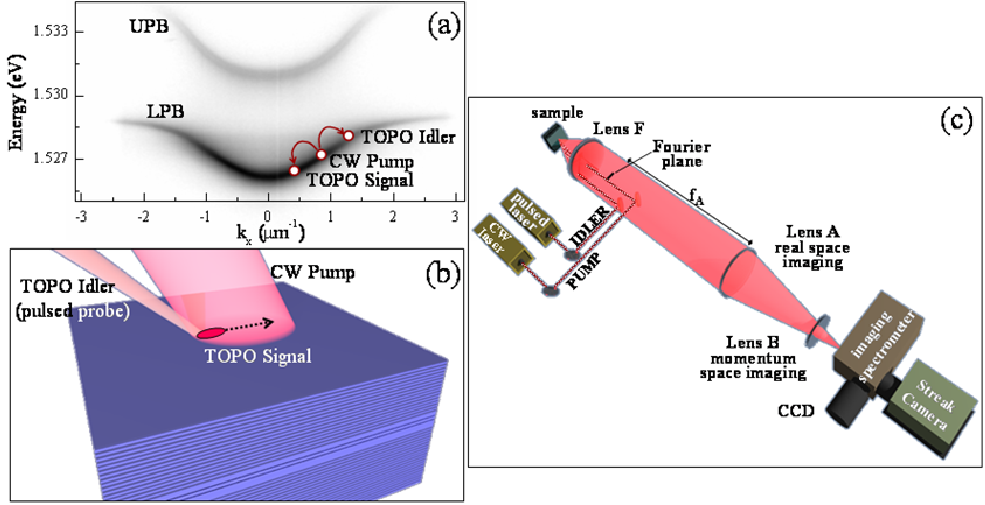}
\caption{(a) Dispersion obtained from the PL emission of the investigated microcavity under
non-resonant excitation at $\delta=0$. In the LPB, a sketch of the
TOPO configuration is presented. The cw pump and pulsed idler arrive
at the sample with angles of $10^{\circ}$ and $16^{\circ}$ respectively,
giving rise to a signal state expected at $4^{\circ}$ from the growth
direction (normal to the surface). (b) Sketch of the pump and idler laser spots on the sample.
(c) Scheme of the excitation and detection set-up. $f_A$ indicates the focal length of lens A.}

\label{FigTOPOsketch}
\end{figure}
%***************************************************************************

Our experimental configuration is based on a continuous replenishing
of the polariton fluid from a higher-lying state driven
by an external cw-laser in a configuration of a triggered optical parametric oscillator (TOPO)
 in the lower polariton branch, based on the OPO physics. Figure~\ref{FigTOPOsketch}~(a)
shows a sketch of the excitation conditions in this configuration.
The sample employed for the experiments described in this section
is a $\lambda/2$-GaAs/AlGaAs microcavity with a single wide QW in
the antinode of the electromagnetic field in the center of the cavity, grown at \emph{Laboratoire de Photonique et de Nanostructures} in France~\cite{Perrin2005b}.
%The sample was grown at the \emph{Laboratoire de Photonique et de
%Nanostructures} (CNRS, France). More details about this sample can be found in Sec.~\ref{sub:LPNMicrocavity}.
The excitation
conditions depicted in figure~\ref{FigTOPOsketch}~(a), are plotted on top of
the PL dispersion under low power, non-resonant excitation, at 10~K
(temperature at which all experiments in this section have been done) and at $\delta=0$.

The way the OPO works is the following. Under cw pumping, a large
polariton population is created at a LPB state (the \emph{pump} state) with energy $E_{P}$ and in-plane momentum $k_{P}$. $E_{P}$ can
be selected by tuning the laser energy, while the in-plane momentum
$k_{P}$ is established by the angle of incidence $\theta_P$ of the CW-pump beam
on the sample {[}$k_{P}=(\omega/c)sin\theta_P$, where $c$ is the speed of light and $\omega$ the frequency of the optical field].
Polariton pair-scattering processes are possible to the signal 
and the idler states, as long as the phase matching conditions between
pump, signal and idler are fulfilled:

%++++++++++++++++++++++++++++++++++++++++++++++++++++++++++++++++++++++++++++
\begin{equation}
2E_{P}=E_{S}+E_{I},
\label{eq:PhaseMatchingEnergy}
\end{equation}
%++++++++++++++++++++++++++++++++++++++++++++++++++++++++++++++++++++++++++++
\begin{equation}
2\boldsymbol{k_{P}}=\boldsymbol{k_{S}}+\boldsymbol{k_{I}}.
\label{eq:PhaseMatchingMoment}
\end{equation}
%++++++++++++++++++++++++++++++++++++++++++++++++++++++++++++++++++++++++++++

If the pump population is large enough, pair scattering is spontaneously
stimulated to the signal at $k_{S}=0$~\cite{Whittaker2001} (i.e. $\boldsymbol{k_{I}}=2\times\boldsymbol{k_{P}}$),
which becomes macroscopically occupied in a well defined quantum state~\cite{Stevenson2000,Savvidis2000,Whittaker2001,Ciuti2001}.
However, if a polariton population at an idler state such that $\boldsymbol{k_{I}}\neq2\times\boldsymbol{k_{P}}$
is created by a cw probe while the cw excitation of the pump state is on, pair scattering
processes will be stimulated to a signal state predetermined by equations~(\ref{eq:PhaseMatchingEnergy}-\ref{eq:PhaseMatchingMoment})~\cite{Sanvitto2005b}.
Thus, the momentum and energy of a signal-polariton population can
be arbitrarily prepared with the proper selection of pump and idler
energies and incidence angles. This configuration is called Optical
Parametric Amplifier (OPA), and can be easily
achieved for a range of pump, signal and idler states due to the peculiar
dispersion found in semiconductor microcavities and the strong non-linearities
associated to the polariton interaction.

In our experiment, the pump is a cw beam and the probe is a short
(1.5~ps) pulse at the idler state, which just initializes the system
inducing a population in the signal. After the probe pulse has
disappeared, the signal state is left macroscopically occupied, and
final state stimulation of polaritons from the pump to the signal 
carries on for nanoseconds. This novel experimental configuration,
only initialized by the probe pulse, corresponds to a triggered OPO (TOPO), where the self-sustained high occupancy of the signal
is fueled by the continuously replenished polariton
population in the pump state (figure~\ref{FigTOPOsketch}).

With the experimental configuration just described, we can create
polariton states with a well defined non-vanishing momentum at an
energy different from that of the excitation lasers. By making use
of a spectrometer in the detection setup, we can select the emission
from the signal states and filter out the stray light from the excitation
sources. Additionally, the high occupancy of the signal state, which
is an essential precondition for the TOPO process to be activated,
assures that signal polaritons are indeed in a macroscopically
occupied state.% Figure~\ref{FigTOPOsketch}~(b) shows an actual time
%integrated CCD image of the far field PL in the regime of OPA, with
%cw pump and idler excitations.

In the time resolved experiments, the cw pump is focused on the sample
on a spot of~$100\,\mu$m in diameter, while the pulsed idler is
focused with a size of $16\,\mu$m inside the pump spot, as depicted
in figure~\ref{FigTOPOsketch}~(b). Once the TOPO is initialized on
the spot of the sample illuminated by the pulsed laser, the signal
polaritons start to move. In a time given by the polariton lifetime in the cavity
(2-4~ps), polaritons displace to a nearby different point 
and then leak out of the cavity. However, the cw pump continues to
feed the signal state by the TOPO mechanism, as the signal
polaritons move within the pump spot. Once the fluid reaches the pump-spot
edge, the signal dies away. The position of the signal-polariton fluid
is evidenced by the light emitted at the energy of the signal state,
which arises from the photons escaping from that state.

In all experiments presented here (except stated otherwise), the cw pump
will be injected at an angle of $10^{\circ}$ (corresponding to an
in-plane momentum of $\sim1.15$~$\mu$m$^{-1}$). %
This angle is slightly below the \emph{magic angle} ($12^{\circ}$, inflexion
point of the LPB). In this way, the threshold for the spontaneous
OPO of the pump is increased and the emission of the OPO signal at
$k=0$, which could contaminate the emission from the nearby TOPO
signal state, is minimized~\cite{Whittaker2001,Whittaker2005}. Nonetheless, this pump angle is close enough
to the inflexion point so that the phase matching conditions given by equations~(\ref{eq:PhaseMatchingEnergy}-\ref{eq:PhaseMatchingMoment})
for the TOPO are still easily achieved~\cite{Butte2002b}. 

Let us note that polariton-polariton interactions result in appreciable
renormalizations (blueshifts) of the LPB when large populations are
injected in the system. These interactions arise from the exciton
content of polaritons. When preparing the pump
and idler states in the TOPO, the band blueshift must be taken into
account by the fine adjustment of the energy of the cw pump and pulsed idler
so that the phase matching conditions are satisfied in the renormalized dispersion.

The setup used for the detection of the polariton fluids moving across
the sample is shown in figure~\ref{FigTOPOsketch}~(c).
Two configurations can be used for detection. For real-space detection, lens B is
removed from the set-up. Lens A forms an image of the sample's surface
on the entrance slit of the spectrometre attached to the streak camera, enabling the construction
of time resolved images of the PL from the sample's surface at a given emission energy. For the recording of the
far field emission (momentum space), lens B is inserted. In this case the
Fourier plane of lens F is imaged onto the entrance slit of the spectrometre. The time resolution
of the obtained films is about 8~ps, while the total spectral and
spatial resolutions are on the order of 0.2~meV and 4~$\mu$m, respectively.

%***************************************************************************************
\begin{figure}
\centering\includegraphics[clip,width=0.5\textwidth]{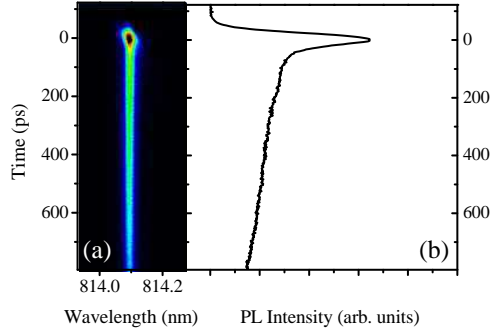}
\caption{Time evolution of the TOPO signal at $k=0$ after the arrival of the
pulsed idler, for pump and idler energies and momenta such that the
phase matching conditions result in a signal at $k=0$ (pump incidence
angle: $12^{\circ}$). (a) Streak-camera image. (b) Time profile, showing a
signal decay-time of $1.1$~ns.}

\label{FigTOPOkZero}
\end{figure}
%***************************************************************************************

The sustainability in time of the TOPO process can be investigated
if the pump and idler beams are set
to in-plane momenta and energies such that the signal state appears
at $k_{S}=0$ {[}$\boldsymbol{k_{I}}=2\times\boldsymbol{k_{P}}$ and $E_{I}=2\times E_{P}-E(k=0)$], following the phase matching
conditions of equations~(\ref{eq:PhaseMatchingEnergy}-\ref{eq:PhaseMatchingMoment}).
In this case the signal polaritons do
not move in space and its lifetime can be measured. Figure~\ref{FigTOPOkZero}~(a)
shows a streak-camera image in its usual configuration (wavelength
\emph{vs} time) of the signal PL at $k=0$. The signal emission is
triggered at the arrival of the idler pulse. The very fast initial
decay is caused by the disappearance of the idler pulse, whose photon
density sets the initial occupation of the signal states. After the
pulse has disappeared the signal is fed by the pump polaritons, showing
a decay of $1.1$~ns {[}figure~\ref{FigTOPOkZero}~(b)]. The details on the parameters which
determine this decay time are discussed in references~\cite{Wouters2007b,Ballarini2009}.

\subsection{Polariton flow in the absence of defects\label{sub:PolaritonNoDefects}}

%*******************************************************************************
\begin{figure}
\centering\includegraphics[clip,width=1\textwidth]{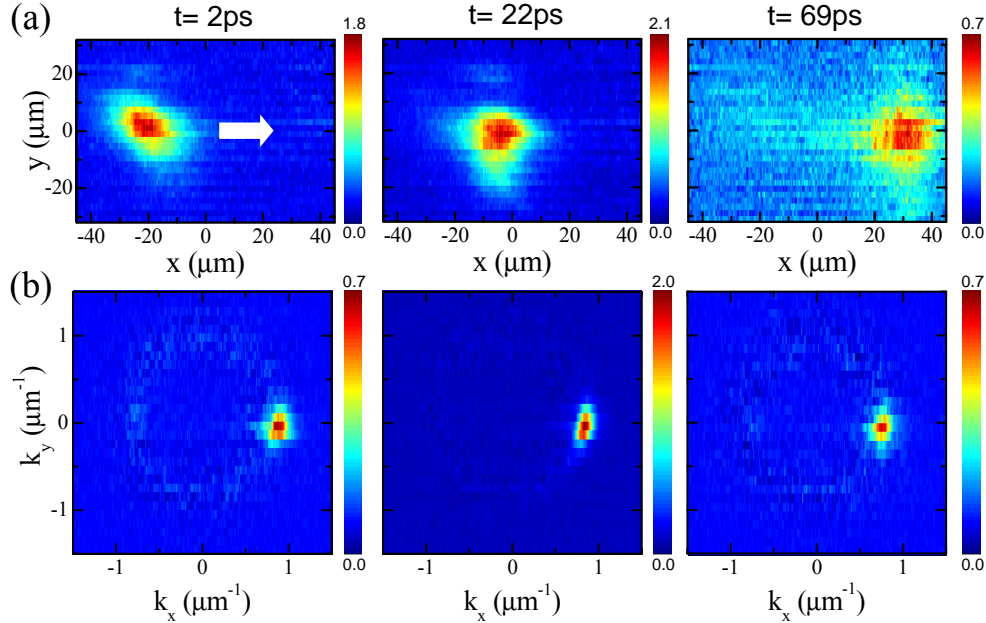}
\caption{(a)~Spectrally selected observation at the TOPO signal energy of a coherent
polariton gas moving at $v_{g}=0.8$~$\mu$m/ps. The images are real
space shots taken at different times after the probe pulse arrival
(t~=~0). (b)~Reciprocal (momentum) space frames recorded at the same
time delays and energy as in (a). The diffusionless
motion and the invariance of the $k$-vector are a clear signature that
polaritons are in a regime showing quantum coherence. The CW pump
power is 43~mW ($10^{\circ}$), while the pulsed probe power is 40~$\mu$W
($16^{\circ}$).}

\label{FigUnperturbedFlow}
\end{figure}
%**********************************************************************************

Figure~\ref{FigUnperturbedFlow}~(a) shows several frames of the real-space
emission at the energy of the signal polaritons, in the TOPO configuration
described in the caption of figure~\ref{FigTOPOsketch}~(a) (i.e. the signal has a non-\emph{zero} momentum), after the
arrival of the pulsed idler. Far field images at the same emission
energy are also displayed for the same time delays {[}figure~\ref{FigUnperturbedFlow}~(b)].
In figure~\ref{FigUnperturbedFlow}, as well as in the rest of the
time sequences presented in this section, the images are obtained by recording
the emission with pump and idler beams impinging upon the sample and
subtracting the emission from just the pump excitation. In this way
only the polaritons populated by parametric processes triggered by
the pulsed idler are recorded. It is readily seen that signal polaritons
freely move across the sample without expanding or interacting with
the surrounding medium until the polariton packet reaches the edge of the area
excited by the cw pump. In $k$-space the motion is unperturbed and
the total polariton momentum is preserved, with a value of $k_{p}=0.85$~$\mu$m$^{-1}$,
without any appreciable spreading.

A detailed analysis of real-space films shows that the polariton fluid
moves at a constant speed. Its group velocity is $v_{g}=(0.8\pm0.2)$~$\mu$m/ps.
We can compare this value with the velocity associated to the observed
momentum in the $k$-space images. For particles described by a parabolic dispersion, the group velocity and the momentum
of the center of mass of the fluid are given by
%++++++++++++++++++++++++++++++++++++++++++++++++++++++++++++++++++++++++++++++++++++
\begin{equation}
v_{g}=\frac{\hbar k_{p}}{m_{p}},
\label{eq:VeloFromMomentum}\end{equation}
%++++++++++++++++++++++++++++++++++++++++++++++++++++++++++++++++++++++++++++++++++++
 where $m_{p}$ is the polariton mass, which can be obtained
from fitting the bottom of the
LPB observed under non-resonant
excitation [figure~\ref{FigTOPOsketch}~(a)] to a parabola 
($m_{p}=\frac{1}{\hbar^{2}}\frac{\partial^{2}E}{\partial k_{p}^{2}}$).
For the conditions of $\delta=0$ considered here a LPB polariton
mass of $1.89\times10^{-4}m_e$ is obtained.

\begin{figure}
\centering\includegraphics[clip,width=0.5\textwidth]{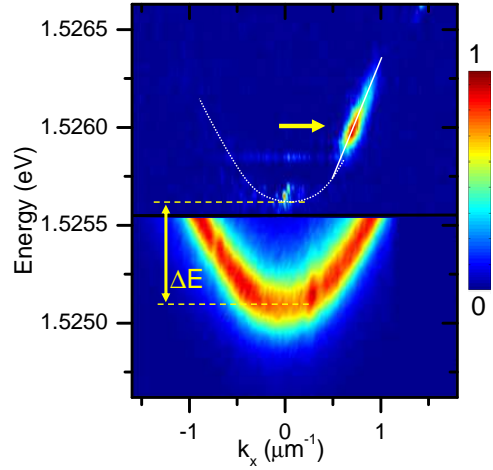}
\caption{ Upper panel: PL intensity as a function of energy and momentum of the emission
for $k_{y}=0$, at short time (12~ps) after the arrival of the pulsed
laser in the TOPO configuration. The white lines are a linear and parabolic
fit to the observed luminescence. Lower panel: polariton dispersion under non-resonant
low power excitation.}

\label{FigResonantDispersion}
\end{figure}

In this case, with the value of $k_{p}$ obtained from figure~\ref{FigUnperturbedFlow}~(b),
a polariton group velocity of $v_{g}=(0.6\pm0.3)$~$\mu$m/ps is
calculated. This value is close to that directly measured
from the movement in real space. However, the value of 
the velocity of the polariton packet should not be given by equation~(\ref{eq:VeloFromMomentum}), which assumes a parabolic dispersion, but by
the actual slope of the renormalized dispersion under 
TOPO conditions:
%++++++++++++++++++++++++++++++++++++++++++++++++++++++++++++++++++++++++++++++++++++
\begin{equation}
v_{p}=\frac{1}{\hbar}\frac{\partial E}{\partial k_{p}}.
\label{eq:velocitydispersion}\end{equation}
%++++++++++++++++++++++++++++++++++++++++++++++++++++++++++++++++++++++++++++++++++++
The upper panel of figure~\ref{FigResonantDispersion}
shows the LPB dispersion around $k_{x}=0$ ($k_{y}$ is kept 0), as
obtained in the TOPO regime at short times (12~ps) after the arrival of the
idler pulse. The emission is spatially integrated
over all the excitation spot, but as shown in figure~\ref{FigUnperturbedFlow},
the PL characteristics do not significantly change along the
trajectory of the fluid. Note that the image is also obtained by recording
the emission in the TOPO (cw pump plus pulsed idler) and subtracting
that caused by the cw pump only.

%*********************************************************************************
%*********************************************************************************

A clear linearization of the emission around the signal state can be easily identified,
as depicted by the linear fit shown in white. For comparison, the lower panel depicts the standard parabolic dispersion
obtained on the same spot under non-resonant low power excitation.
A strong blueshift of the emission ($\Delta E=0.5$ meV) in the TOPO conditions is also evidenced. Both the linearization and the renormalization
arise from the non-linear response of the system due to the dominant polariton-polariton interactions
that dress the state in such a high density phase. This type of interactions has been predicted to lead to the appearance of
superfluidity~\cite{Carusotto2004} and solitons~\cite{Yulin2008,Larionova2008,Egorov2009} in polariton gases.
The linearized dispersion resembles the Bogoliubov
dispersion for elementary excitations of a superfluid. In that case,
linearization leads to suppression of weak scattering and therefore to
motion without dissipation. In ours, the situation is more complicated, as the observed dispersion reflects the
dynamics of the whole system (pump, signal and idler states) as shown in the calculations based on a non-linear Schr\"{o}dinger equation developed in reference~\cite{Amo2009}.

With these premises, we can consider the signal state as a macroscopically degenerate polariton
fluid moving on top of a linearized dispersion. In this case we can calculate the fluid speed associated to
the slope of the dispersion shown in figure~\ref{FigResonantDispersion} making use of equation~(\ref{eq:velocitydispersion}).
For $k_{p}=0.85$~$\mu$m$^{-1}$ (where the dispersion is linear)
we obtain a velocity of $v_{g}=(1.7\pm0.5)$$\mu$m/ps, comparable
to the value of $(0.8\pm0.2)$~$\mu$m/ps directly obtained from
the real space film in figure~\ref{FigUnperturbedFlow}~(b). Even though one would expect a closer agreement between the observed $v_{g}$ and the slope of the dispersion, the actual relationship between $k_{p}$, $v_{g}$ and the dispersion needs to be established via the detailed study of the non-linear Schr\"{o}dinger equation in this system~\cite{Amo2009}.

%*********************************************************************************
\begin{figure}
\centering\includegraphics[clip,width=0.5\textwidth]{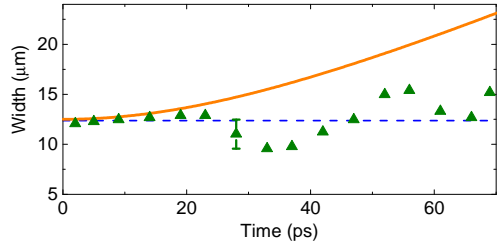}
\caption{Solid triangles: Gaussian width of the polariton packet in the \emph{y}-direction
extracted from the real space images of figure~\ref{FigUnperturbedFlow}~(a).
Solid orange line: calculated wavepacket diffusion for a parabolic dispersion [equation~(\ref{eq:DiffusionParabolic})].
Blue dashed line: expected
wavepacket diffusion in a linear dispersion.}

\label{FigDiffusionWavepacket}
\end{figure}
%*********************************************************************************

\subsection{Polariton diffusion}

A direct consequence of the signal polaritons moving on top of a linearized
section of the dispersion as that shown in figure~\ref{FigResonantDispersion},
is that the signal polariton wavepacket does not spread. A wavepacket
of non-interacting particles on a parabolic dispersion is subject
to real- and momentum-space expansion due to the effects of the uncertainty
principle (non-commutability of space and momentum operators).
Under such circumstances, if the considered wavepacket has a size
(full width at half maximum) of $\Delta_{0}$ at $t=0$, the time evolution of its Gaussian
width $\Delta$ is given by~\cite{Cohen1977}:
%+++++++++++++++++++++++++++++++++++++++++++++++++++++++++++++++++++++++
\begin{equation}
\Delta=\sqrt{\Delta_{0}^{2}+\left(\frac{2\hbar t}{m_{p}\Delta_{0}}\right)^{2}}\label{eq:DiffusionParabolic}\end{equation}
%+++++++++++++++++++++++++++++++++++++++++++++++++++++++++++++++++++++++
However if the wavepacket lives on a linear dispersion, as is the
case of the signal polariton fluid, no expansion at all is expected,
 neither in real nor in momentum space~\cite{Eiermann2003}. Figure~\ref{FigDiffusionWavepacket}
depicts in solid triangles the Gaussian width, in the $y$ direction, of
the signal polariton fluid shown in figure~\ref{FigUnperturbedFlow}~(a)
as it traverses the excitation spot. No apparent diffusion of the
polariton packet is observed. The orange solid line displays the time
evolution of the width of the wavepacket if the polariton fluid would
be characterized by the parabolic dispersion [equation~(\ref{eq:DiffusionParabolic})].

A similar behavior is expected in reciprocal (momentum) space. Indeed,
figure~\ref{FigUnperturbedFlow}~(b) shows no apparent diffusion of
the momentum of the polariton wavepacket.

%***********************************************************************************
\begin{figure}
\centering\includegraphics[clip,width=0.9\textwidth]{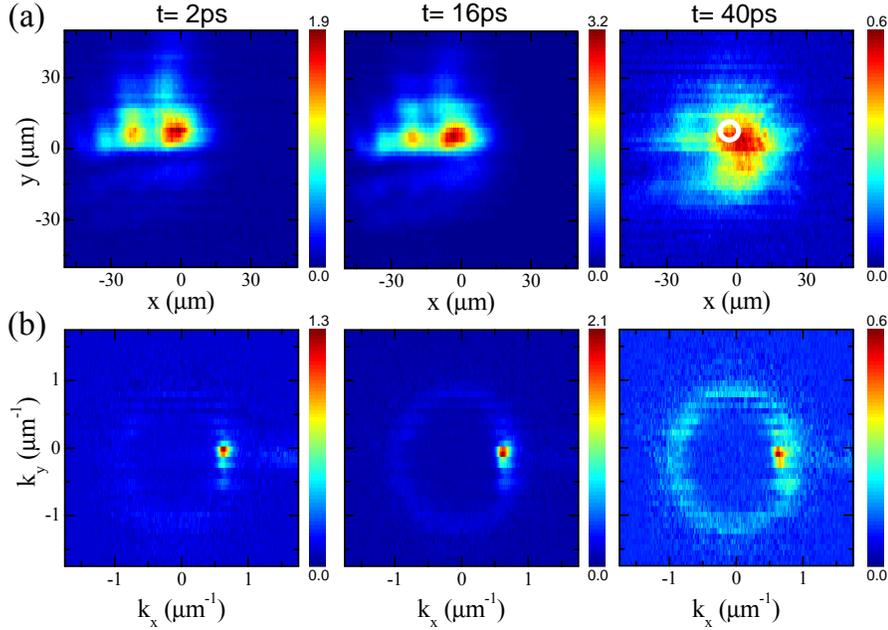}
\caption{Real-space (a) and momentum-space (b) images, of a slow polariton
fluid getting trapped in a shallow potential. The images are obtained
at low pump power (19~mW, $10^{\circ}$). The idler is set at $16^{\circ}$
at $150$~$\mu$W. In real space the fluid slowly moves 
to the right. The white circle at 40~ps depicts the position
of the fluid at $t=2$~ps. At $t\gtrsim40$~ps the fluid gets localized
in the shallow trap.}

\label{FigIncoherentPolariton}
\end{figure}
%***********************************************************************************

It is interesting to compare the behavior of a characteristic polariton
quantum fluid as that depicted in figures~\ref{FigUnperturbedFlow} and~\ref{FigDiffusionWavepacket},
with a polariton packet in the incoherent regime. Figure~\ref{FigIncoherentPolariton}
shows the time evolution in real (a) and momentum (b) space at the
energy of the signal polaritons for a low power (19~mW) cw pump excitation.
The emission from the signal polaritons is still triggered by the
arrival of the pulsed idler. In this case, the fluid starts to move slowly
with a well defined momentum (first and second columns) but gets trapped
in a wide shallow potential region of the sample (third column). In momentum space,
a Rayleigh ring forms, which is a direct evidence of
the signal state not being in a well defined quantum state, but scattering into
a plethora of incoherent states with different momentum. Still, for
the first $\sim20$~ps the fluid has a favored momentum, which reflects
the slow movement of the ensemble at those early times.

\subsection{Scattering with small point-like defects}

In order to explore in detail the superfluid character of the signal (and pump)
polariton fluids, we will present results on
the interaction of these fluids with localized potential barriers. Localized potential barriers are present in the
microcavity in the form of photonic or excitonic defects. In the sample
under study, which is of a very high crystalline quality~\cite{Perrin2005b}, scattering
centers are present with an approximate density of $0.01$~$\mu$m$^{-2}$.
The interaction of a polariton fluid with such defects
on the sample may reveal its quantum nature and superfluid properties,
analogously to the criteria employed to call for superfluid character in atomic BEC~\cite{Onofrio2000,Fort2005}

%*******************************************************************************************
\begin{figure}
\centering\includegraphics[clip,width=1\textwidth]{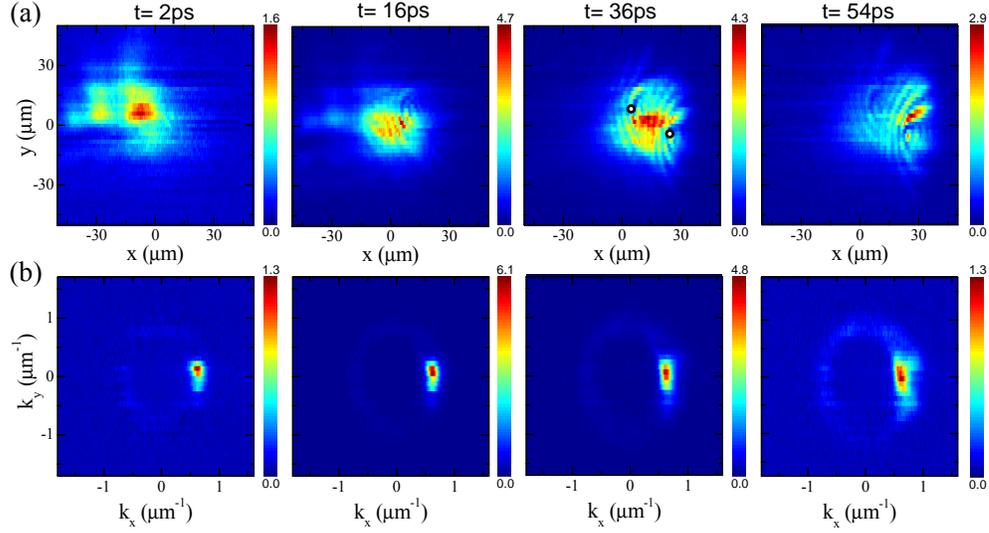}
\caption{(a) Signal polariton movement in real space at different times in
the presence of two small defects (marked with white dots in the third panel).
The observed waves reflect the
local change in density of the \emph{pump} polaritons, which do interact with the defects (see figure~\ref{FigRealSpaceTopoSketch}).
(b) Corresponding momentum space images. The pump (idler) power is 21~mW (150~$\mu$W) and
the angle of incidence is $10^{\circ}$($16^{\circ}$).}

\label{FigSuperfluidMovement}
\end{figure}
%*******************************************************************************************

Figure~\ref{FigSuperfluidMovement}~(a) shows images obtained in the
near field of a TOPO polariton-fluid in the same spot of the sample as in figure~\ref{FigIncoherentPolariton}. In this case
the pump intensity was increased to 21~mW, which is enough for the pump field to induce a renormalization
of the polariton potential so as to make disappear the shallow potential which led to 
the localization of the wave packet in figure~\ref{FigIncoherentPolariton}.
In the present conditions, the packet is able to flow a long distance (40~$\mu$m), encountering
two deep localized defects in its trajectory {[}marked with white points in the third panel of figure~\ref{FigSuperfluidMovement}~(a)]. In the course of its propagation, the
signal shows unambiguous signs of passing through the defects.
However, it clearly maintains its cohesion in this process. This is
most strikingly observed in momentum space [figure~\ref{FigSuperfluidMovement}~(b)], where the
signal is left completely unaltered until the very end of the trajectory,
when the single-state occupancy starts spreading as the signal dies by
moving off the edge of the pump laser region.

%**************************************************************************
\begin{figure}
\centering\includegraphics[clip,width=0.5\textwidth]{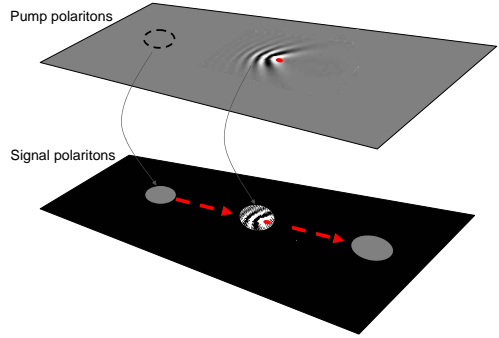}
\caption{Sketch of the TOPO in real space. The observation of the signal polariton-fluid
is represented by the circles running from left to right on the black
background. We are able to detect this motion thanks to the continuous
feeding from the pump polaritons which are represented by the gray
plane. The supersonic regime of the pump polaritons is evidenced by
the presence of density waves around the defect.
The change in density of the pump polaritons is projected into the
signal polaritons which, instead, move through the defect. The red
point shows the position of the defect.}

\label{FigRealSpaceTopoSketch}
\end{figure}
%**************************************************************************

The images reflect the addition of two different contributions: ($i$) the
pump polaritons (extended in an area of $\sim8\times10^{3}$~$\mu$m$^{2}$)
which constantly feed the signal polariton, and ($ii$) the motion of the
signal polaritons by themselves, which pass through the defects. Figure~\ref{FigRealSpaceTopoSketch}
illustrates how these two contributions are detected at the signal
polariton energy. The fringes observed around the defects appear due
to the local change in density of the pump polaritons, which is reflected
in the structure of the signal.

The pump polaritons are injected in a coherent state, at high energies,
high density and with high $k$-vector. At such high momentum pump polaritons interact with the
defects giving rise to interference waves in the upstream direction. This interaction is analogous to that
observed in an atomic condensate flowing at a speed higher than its sound speed against a 
potential barrier~\cite{Carusotto2006}. The same prediction has been established~\cite{Ciuti2005} and recently
observed~\cite{Amo2008} for the case of polaritons under resonant excitation by a single pump field. By contrast,
when the flow speed is lower than the sound speed (subsonic), a superfluid in the
macroscopically occupied condensate is expected~\cite{Onofrio2000,Ciuti2005}.

It is important to note that the visibility of these waves does not
imply that the signal polaritons are also in such a \emph{perturbed}
regime. On the contrary, the signal polaritons (which are at lower
energy and wavevector) seem to flow unperturbed when passing through
the defects. In fact, figure~\ref{FigSuperfluidMovement}~(b) shows that
the signal polaritons present a very narrow and well defined momentum
(evidencing its quantum-state nature) that hardly changes when passing
through the defects. This non scattering behaviour of signal polaritons evidences their superfluid character: excitations in the droplet are inhibited while it passes
through the obstacle, equivalent to the subsonic situation described above.

The key element to understand the coherent propagation of the polariton droplet revealed in figure~\ref{FigSuperfluidMovement},
is the linear dispersion shown in figure~\ref{FigResonantDispersion}. All particles in the 
packet remain in phase and at the same group velocity, preventing its diffusion both in real
and momentum spaces, even when flowing through localized defects.  Other examples of coherent motion through defects of different sizes can be found in~\cite{Amo2009}. As recently shown theoretically, solitonic solutions are expected in semiconductor microcavities under OPO operation in the presence of a linearized dispersion~\cite{Egorov2009}. However these solutions are restricted to one-dimensional architectures, for very precise values of the pump momentum, with solitons which are extended over the ensemble of signal, pump and idler modes. Contrary to that, our results show wavepacket propagation in two-dimension, for the signal state, for a number of pump momentum realizations (not shown).

%% file: conclusions.tex
\section{Summary and future perspectives}

In the first part of this tutorial we have presented experimental results demonstrating
the capability to optically control in a short time-scale ($<2$~ps)
the distributions of carriers in QWs. In particular, we have
seen that the arrival of a short light pulse results in a sudden warming
of a pre-excited electron-hole plasma. The exciton distributions,
which are in thermal equilibrium with the electron-hole plasma, are
also abruptly warmed-up by the arrival of the pulse, resulting in an
ultrafast dip in the \emph{hh}-exciton luminescence.

We have introduced a model that quantitatively reproduces the observed
luminescence quenching and can be used to obtain the relative increase
in carrier/exciton temperature induced by the arrival of the pulse,
by measuring the magnitude of the dip in the \emph{hh}-exciton PL.

This model has been used to study the polariton relaxation from the \emph{reservoir}
states to the bottom of the LPB in semiconductor microcavities. Understanding the
mechanisms of this relaxation is of great importance to improve the design of
microcavites with enhanced conditions for the spontaneous condensation of polaritons.
In the studied microcavities, under non-resonant excitation, condensation could not be observed
at any detuning. By contrast, the transition from the strong to the weak coupling regimes
and the simultaneous onset of lasing is observed, and reported in detail, when the excitation density is
increased above a certain threshold.

In the final part of the tutorial we have presented an experimental configuration based 
on the optical parametric oscillator regime, which
allows for the creation and detection of polariton quantum fluids
with a non-\emph{zero} momentum. The high pump-polariton density renormalizes
the LPB and changes its shape, due to the strong polariton interactions.
The generated signal fluid lives on top of a linearized dispersion and shows no
diffusion in both real and momentum space. Furthermore, the signal packet is able to traverse
defects encountered in its flow path without scattering. These observations
are compatible with the description of the signal fluid as a superfluid.

The phenomenology associated to bosonic quantum fluids has just been
uncovered in atomic condensates in the past decade, and the subtle links
between condensation and superfluidity have been only recently established~\cite{Legget1999}.
 For instance, in the regime
of quantum reflections, very rich effects have been recently observed~\cite{Pasquini2004,Pasquini2006}
and studied theoretically~\cite{Scott2005,Scott2007,Martin2007}
in atomic BECs interacting with surfaces. There is plenty of room for
the exploration of these effects in semiconductor microcavities, with
the advantage of the simple implementation of barriers and defects
with on-demand shapes and sizes by use of lithographic techniques~\cite{Lai2007b,Kim2007}.

Faraday waves~\cite{Engels2007} and Josephson oscillations~\cite{Raghavan1999,Albiez2005,Levy2007}
already observed in atomic condensates are some of the phenomena which could be directly
explored in semiconductor microcavities~\cite{Sarchi2008}, for instance by use
of surface acoustic waves~\cite{deLima2006}
or by creating two nearby-connected polariton condensates~\cite{Shelykh2008}.

Additionally, the dynamic (rather than thermal) equilibrium in
a semiconductor microcavity caused by the constant introduction and
escape of polaritons in the system, presents important differences to the
atomic case. These differences result in a rich phenomenology unaccessible
with the study of atomic condensates. First studies in this direction are the 
spontaneous formation of full~\cite{Lagoudakis2008,Keeling2008} and half vortices~\cite{Rubo2007,Lagoudakis2009} without stirring of the condensate, and
the appearance of a diffusive Goldstone mode in the OPO regime~\cite{Wouters2007b,Ballarini2009}.

In semiconductor microcavities, condensates and fluids are easily
manipulated with the excitation laser-beams, and are currently created
with standard liquid-He cryogenic techniques, at 5-20~K. Furthermore,
wide bandgap systems, such as those based on GaN have already shown
very promising results~\cite{Christopoulos2007,Christmann2008} on the prospective
creation of Bose-Einstein polariton condensates at room temperature.